\documentclass[prb,twocolumn,showpacs,superscriptaddress,amsmath,floatfix]{revtex4-1}
\usepackage{amssymb}
\usepackage{citesort}
\usepackage{dcolumn}
\usepackage{graphics,psfrag}

\begin{document}
\title{Coulomb enhancement of superconducting  pair-pair correlations in a $\frac{3}{4}$-filled
  model for $\kappa$-(BEDT-TTF)$_2$X}
\author{W. Wasanthi De Silva}
\affiliation{Department of Physics and Astronomy and HPC$^2$ Center for 
Computational Sciences, Mississippi State University, Mississippi State MS 39762}
\author{N. Gomes}
\affiliation{Department of Physics, University of Arizona, Tucson, AZ 85721}
\author{S. Mazumdar}
\affiliation{Department of Physics, University of Arizona, Tucson, AZ 85721}
\author{R. T. Clay}
\affiliation{Department of Physics and Astronomy and HPC$^2$ Center for 
Computational Sciences, Mississippi State University, Mississippi State MS 39762}

\date{\today}
\begin{abstract}
We present the results of precise correlated-electron calculations on
the monomer lattices of the organic charge-transfer solids
$\kappa$-(BEDT-TTF)$_2$X for 32 and 64 molecular sites. Our
calculations are for band parameters corresponding to X =
Cu[N(CN)$_2$]Cl and Cu$_2$(CN)$_3$, which are semiconducting
antiferromagnetic and quantum spin liquid, respectively, at ambient
pressure.  We have performed our calculations for variable electron
densities $\rho$ per BEDT-TTF molecule, with $\rho$ ranging from 1 to
2. We find that $d$-wave superconducting pair-pair correlations are
enhanced by electron-electron interactions only for a narrow carrier
concentration about $\rho=1.5$, which is precisely the carrier
concentration where superconductivity in the charge-transfer solids
occurs. Our results indicate that the enhancement in pair-pair
correlations is not related to antiferromagnetic order, but to a
proximate hidden spin-singlet state that manifests itself as a
charge-ordered state in other charge-transfer solids.
Long-range
superconducting order does not appear to be present in the purely
electronic model, suggesting that electron-phonon interactions also
must play a role in a complete theory of superconductivity.
\end{abstract}
\maketitle

\section{Introduction}
\label{sec-introduction}

The family of layered organic superconductors $\kappa$-(BEDT-TTF)$_2$X
(hereafter $\kappa$-(ET)) has attracted strong interest because of its
apparent similarity with the high T$_c$ cuprates. As in the cuprates,
superconductivity (SC) in $\kappa$-ET is proximate to semiconducting
magnetic states, antiferromagnetic (AFM) or quantum spin liquid
(QSL) \cite{Kanoda11a}. SC in the $\kappa$-(ET)$_2$X as well as
organic charge-transfer solids (CTS) in general is however reached by
application of pressure rather than doping with charge carriers
\cite{Ishiguro}. Thus SC is a consequence of change of one or more
parameters in the Hamiltonian that describes both the semiconducting
and superconducting states, at fixed carrier concentration.  The
$\kappa$-ET lattice is strongly dimerized, with strong intradimer
electron hoppings between ET molecules belonging to the same dimer,
and relatively weaker interdimer hoppings (see
Fig.~\ref{fig-lattice}). Complete CT of one electron to the acceptor
molecule X occurs from each dimer, which creates cations
  ET$^{0.5+}$ with 0.5 holes (1.5 electrons) in the highest occupied
  molecular orbital (MO) of each ET molecule. The dimer lattice is
anisotropic triangular, and the magnetic behavior of the family can be
understood within an {\it effective} $\frac{1}{2}$-filled band Hubbard
model, wherein the dimers  ET$^2{+}$ of the ET molecules (rather than the
molecules themselves) constitute the individual sites
\cite{Ishiguro,Kino95a}.  AFM in the strongly anisotropic X =
Cu[N(CN)$_2$]Cl (hereafter $\kappa$-Cl), and QSL behavior in the
nearly isotropic X = Cu$_2$(CN)$_3$ (hereafter $\kappa$-CN) are both
expected within the effective model \cite{Kanoda11a}.

In the context of the cuprates, there exists a large body of
theoretical literature claiming that SC occurs within the Hubbard
model \cite{Anderson87a,Scalapino95a,Anderson04a} for bandfilling
slightly away from $\frac{1}{2}$ but no consensus has yet been reached
on this matter. In analogy to these theories, mean-field and dynamic
mean-field theories of $\kappa$-(ET)$_2$X have proposed that SC also
occurs within the exactly $\frac{1}{2}$-filled band Hubbard model on
an anisotropic triangular lattice, for a range of Hubbard $U$ and
anisotropy
\cite{Vojta99a,Schmalian98a,Kino98a,Kondo98a,Powell05a,Gan05a,Sahebsara06a,Watanabe06a,Kyung06a,Hebert15a}.
A necessary condition for SC within any interacting electrons model
however is that interactions must enhance superconducting pair-pair
correlations relative to the noninteracting model. Precise numerical
calculations within the $\frac{1}{2}$-filled band Hubbard model on
triangular lattices have shown that pair-pair correlations decrease
with Hubbard $U$ for all anisotropy
\cite{Clay08a,Tocchio09a,Dayal12a,Gomes13a}, thus indicating the need
for going beyond the effective 1/2-filled band model in our search for
the the mechanism of SC in $\kappa$-(ET)$_2$X.

Recently an ubiquitous charge-ordered (CO) phase of unknown origin has
been discovered in the cuprates that competes with both AFM and SC
\cite{Chang12a,Wu11a,Blanco-Canosa14a,Hucker14a,Comin14a,SilvaNeto14a,SilvaNeto15a,Wu13a}.
The CO and SC orders have similar energy scales, and some
investigators have suggested that
``CO and SC appear as joint instabilities of the same normal state'' \cite{Wu13a}.
Given that SC proximate to CO is also seen in
(ET)$_2$X with crystal structures different from $\kappa$
\cite{Andres05a,Bangura05a,Tamura06a,Shikama12a}, it is natural to ask
whether the mechanisms of SC in all (ET)$_2$X are actually related,
and whether even in the $\kappa$-ET family there exists a hidden
competing CO state.  Uniquely for the case of carrier density
precisely 0.5 or 1.5 per site, we have shown that there can indeed
occur a {\it spin-paired} CO state, a paired-electron crystal (PEC),
in frustrated lattices \cite{Li10a,Dayal11a}. The PEC is a Wigner
crystal of spin-bonded pairs, rather than of single electrons
\cite{Moulopoulos92a}. In the $\kappa$ materials, the PEC would
require {\it unequal charge densities} on the monomer molecules that
constitute each ET dimer, with {\it inter}dimer spin-pairing between
monomers with large charge densities \cite{Li10a,Dayal11a}.  Other
groups have also proposed related theories of fluctuating intradimer
charge disproportionation \cite{Gomi10a,Naka10a,Hotta10a}, in order to
explain the peculiar dielectric responses of $\kappa$-CN
\cite{Abdel-Jawad10a} and $\kappa$-Cl \cite{Lunkenheimer12a}.  While
recent experiments have shown the absence of static CO in both
materials \cite{Sedlmeier12a}, and ascribe the electrodynamic response
of $\kappa$-CN to coupling of ET cations to anions \cite{Dressel16a}, dynamic
fluctuating CO is not precluded
\cite{Sedlmeier12a,Girlando14a,Yamamoto}.  Such a fluctuating CO may
perhaps explain the low temperature ``6 K transition'' in $\kappa$-CN
that remains unexplained
\cite{Yamashita08a,Yamashita09a,Manna10a,Poirier12a}.  Experimentally,
static PEC has been observed in the $\kappa$ compound X =
Hg(SCN)$_2$Cl \cite{Drichko14a}, while pressure-induced transition
from a dimer AFM to an intradimer charge segregated state has been
observed in the compound $\beta^{\prime}$-(BEDT-TTF)$_2$-ICl$_2$
\cite{Hashimoto15a}, which like $\kappa$-(ET) has a dimerized lattice
structure.

 We have recently proposed that SC in the (ET)$_2$X evolves from a
 {\it paired-electron liquid} (PEL), which can be thought of as a
 destabilized PEC \cite{Gomes15a} (thus a static PEC is not a 
requirement for the PEL or SC.) Our conclusion is based on precise
 numerical calculations of pair-pair correlations within the Hubbard
 model on anisotropic triangular lattices, for variable carrier
 densities $\rho$ per site. We showed that in each case the pair-pair
 correlations are enhanced relative to the noninteracting limit
 uniquely for $\rho \simeq 0.5$. At all other $\rho$ ($0 \leq \rho
 \leq 1$) pair-pair correlations are suppressed by the Hubbard $U$.
 Although the PEL does not have true long-range superconducting
 correlations, to the best of our knowledge this was the first
 demonstration of enhancement of pair-pair correlations within the
 Hubbard model. The fundamental theoretical picture that emerges is
 related to an earlier proposition that SC can evolve from the paired
 Wigner crystal in the electron gas at intermediate densities, upon
 weak doping \cite{Moulopoulos92a}.  Our work has extended this idea
 of evolution from a a paired crystal to a paired liquid to the case
 of a real lattice. Both paired states on the lattice are however
 unique to a particular carrier density.

The anisotropic triangular lattice lacks the strong dimerization of
$\kappa$ lattice, and the possibility of mapping to the effective
$\rho=1$ model \cite{Kino95a} thus does not exist in this
case. Whether or not Coulomb-induced enhancement of pair-pair
correlations occurs in the {\it realistic} $\kappa$ lattice, {\it also
  uniquely for the same carrier concentration $\rho$,} as well as the
symmetry of the superconducting order parameter, if any, are of strong
interest.  We report here the results of such calculations of
spin-spin and pair-pair correlations on the actual $\kappa$ lattice,
for realistic band parameters appropriate for $\kappa$-Cl and
$\kappa$-CN.  As before \cite{Gomes15a}, we perform these calculations
for variable $\rho$.  We demonstrate PEL formation on the $\kappa$
lattices uniquely at or near $\frac{3}{4}$ filling with electrons,
giving further credence to our proposal\cite{Gomes15a} that there
occurs an effective electron-electron (e-e) attraction selectively at
this $\rho$.

The majority of computational studies of correlated models
for $\kappa$-ET have focused on
  the effective $\frac{1}{2}$-filled band model. 
 For the monomer lattice exact diagonalizations within 16-site
clusters within the Hubbard and extended Hubbard models have been
performed  \cite{Clay05c,Gomi10a,Gomi13a,Gomi16a}, as well
as dynamical mean-field (DMFT) studies \cite{Ferber14a}
To our knowledge, the present results report the first
direct calculations of   pair-pair correlations on large lattices.
System sizes available to exact diagonalization are able to
demonstrate spin correlations consistent with AFM order between
dimers, but are not large enough for measurement of pair-pair
correlations between non-overlapping pairs.

The outline of the paper is as follows: in Section \ref{sec-lattice}
we describe the theoretical model, the lattices and the computational
methods we use; in Section \ref{sec-results} we present our
computational results for the spin structure factor and pair-pair
correlations in the ground state; and in Section \ref{sec-discussion}
we discuss our results in relationship to the current experimental
data on $\kappa$-ET, as well as implications for theories of
correlated-electron SC in general.

\section{Theoretical model, lattice, parameters, and methods}
\label{sec-lattice}

\begin{figure}[tb]
  \centerline{\resizebox{2.6in}{!}{\includegraphics{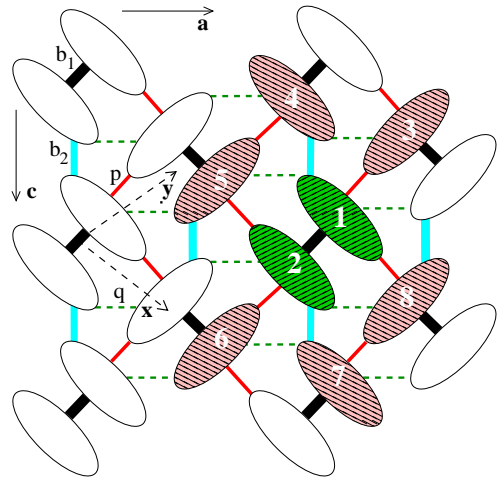}}}
  \caption{(color online) Lattice structure of $\kappa$-(ET)$_2$X,
    showing individual monomer molecules. In order of
    decreasing magnitude, $b_1$, $b_2$, $p$, and $q$ label the
    intermolecular hopping integrals. Superconducting pairs are
    constructed from the shaded molecules numbered 1$\cdots$8 as
    detailed in Section \ref{sec-pair}. The {\bf a} and {\bf c}
    crystal axes for $\kappa$-Cl are indicated; for $\kappa$-CN the
    corresponding axes are conventionally labeled {\bf b} and {\bf
      c}. The {\bf x} and {\bf y} axes for the effective dimer model
    are indicated by dashed arrows.}
  \label{fig-lattice}
\end{figure}
As a minimal model for the electronic properties of
the conducting layer of BEDT-TTF molecules in
$\kappa$-ET, we consider the Hubbard Hamiltonian,
\begin{equation}
  H = \sum_{\langle ij\rangle,\sigma}t_{ij}(c^\dagger_{i,\sigma}c_{j,\sigma}+H.c.)
  + U\sum_i n_{i,\uparrow}n_{i,\downarrow}.
  \label{ham}
\end{equation}
In Eq.~\ref{ham}, $c^\dagger_{i,\sigma}$ ($c_{i,\sigma}$) creates
(annihilates) an electron of spin $\sigma$ on the highest MO of a {\it monomer} ET molecule $i$,
$n_{i,\sigma}=c^\dagger_{i,\sigma}c_{i,\sigma}$, and $U$ is the onsite
e-e interaction. The lattice structure\cite{Mori99a} of the
conducting layers in $\kappa$-ET is shown in Fig.~\ref{fig-lattice}.
In order to differentiate our approach from theories emphasizing the
effective $\frac{1}{2}$-filled band picture, in what follows instead
of using the ``bandfilling'' we will present our computational results
as a function of the average electron density per monomer molecule,
$\rho$. As mentioned above, $\rho$ in the (ET)$_2$X family is 1.5.
While our calculations of the spin-spin correlations are for this
density only, we have performed the calculations of superconducting
pair-pair correlations for a wide range of $\rho$, $1\leq\rho\leq2$.  The
motivation behind studying the density dependence of pair-pair
correlations is two-fold. First, this allows us to investigate whether
or not SC is unique to $\rho \simeq 1.5$, which is a necessary
condition if the SC is indeed evolving from a hidden PEC \cite{Gomes15a}.
Secondly, it also allows us to probe carrier density slightly away
from the stoichiometric $\rho=1.5$ in view of recent
experiments\cite{Oike15a} that have suggested that SC can occur in the
$\kappa$-ET system for weak doping away from $\rho=1.5$.

Hopping integrals for $\kappa$-ET have been previously calculated
using the extended H\"uckel \cite{Mori99a} and density-functional
methods \cite{Kandpal09a,Nakamura09a,Koretsune14a}. Each molecule in
the lattice has significant overlaps with six nearest neighbors (see
Fig.~\ref{fig-lattice}) with hopping integrals $t_{b1}$, $t_{b_2}$,
$t_p$, and $t_q$ in order of decreasing magnitude. In the effective
dimer model the two molecules connected by $t_{b1}$ are considered a
single effective site, with the dimers forming an anisotropic
triangular lattice.  The effective hopping integrals along $x$ and $y$
are $t=(t_p+t_q)/2$ and the frustrating hopping integral
$t^\prime\equiv t_{-x+y}=t_{x-y}=t_{b2}/2$. The degree of frustration
in the effective dimer model is then given by the ratio $t^\prime/t =
t_{b_2}/(t_p+t_q)$.
Frustration is weakest in $\kappa$-Cl and strongest in $\kappa$-CN;
these are the systems we consider.

In our calculations we have used the following sets of $t_{ij}$,
($t_{b1}$, $t_{b2}$, $t_p$, $t_q$), given in meV, from Reference
\onlinecite{Koretsune14a}: for $\kappa$-Cl (207, 67, -102, -43), and
for $\kappa$-CN (199, 91, -85, -17). Both sets were
determined\cite{Koretsune14a} from low-temperature crystal data, $T$=5
K in the case of $\kappa$-CN, and $T$=15 K in the case of $\kappa$-Cl.
For these two sets of parameters, the ratio $t^\prime/t$ is 0.46
(0.89) for $\kappa$-Cl ($\kappa$-CN). While different computational
techniques report somewhat different $t^\prime/t$, all have found that
in terms of the effective dimer model $\kappa$-CN is significantly
more frustrated and closer to an isotropic triangular dimer lattice than is
$\kappa$-Cl. It is however not known how the effect of the larger
frustration within the dimer model affects the electronic properties
of the full monomer lattice.

We considered two different periodic lattices with 32 and 64 molecular
sites. The 32 site lattice is four dimers each along the {\bf c} and
{\bf a} directions in Fig.~\ref{fig-lattice}.
The lattice is chosen such that the effective dimer lattice ({along
  the \bf x} and {\bf y} axes of Fig.~\ref{fig-lattice}) is a
4$\times$4 square lattice.
This is possible if the vectors defining the edges of the 32 site
cluster are along the {\bf c} and {\bf y} directions in
Fig.~\ref{fig-lattice}.  The 64 site cluster was constructed in a
similar way and corresponds to an 8$\times$4 lattice in terms of
dimers.
The full lattices are shown in the Supplemental Material \cite{sm}.
On 
both lattices, the $t_{ij}$  parameters \cite{Koretsune14a} for
$\kappa$-Cl and $\kappa$-CN gave different single-particle  Fermi level
degeneracies; the  degeneracy
for $\rho=1.5$ 
is twofold in $\kappa$-Cl and
fourfold in $\kappa$-CN, in agreement with the greater frustration in
the latter.

Conventional quantum Monte Carlo methods cannot be used in the highly
frustrated $\kappa$-ET lattice due to the fermion sign problem. The
two methods we used are the Path Integral Renormalization Group (PIRG)
 \cite{Kashima01b} and Constrained Path Monte Carlo
(CPMC) \cite{Zhang97a}. Both PIRG and CPMC are ground state projector
methods that project out the ground state from an arbitrary initial
wavefunction. In PIRG the projection is done in a finite basis of
Slater determinants, followed by an extrapolation in the energy
variance \cite{Kashima01b}. In CPMC the projection is done using
random walkers constrained by a trial wavefunction
\cite{Zhang97a}. Here we have used the $U=0$ wavefunction for the
constraint. We have extensively benchmarked calculations of
superconducting pair-pair between these two methods in previous work
on the anisotropic triangular Hubbard model both at $\frac{1}{2}$-filling
\cite{Dayal12a} as well as the complete density range \cite{Gomes15a}.
While PIRG can be considered exact provided large enough basis sets
are used and the projection is done with care, the constraining
wavefunction in CPMC does introduce a systematic error. In our
previous work, we found that CPMC results for pairing correlations
agreed well with PIRG for small to intermediate $U$, provided use of
CPMC was restricted to systems  which in the noninteracting $U=0$ limit have nondegenerate
closed-shell Fermi level occupancies \cite{Zhang97a,Gomes15a}.

\section{Results}
\label{sec-results}
We performed PIRG calculations for the 32-site lattice over the full
density range $1<\rho<2$ and for $\rho=1.5$ for the 64-site lattice.
  For the other densities of the 64 site lattice with nondegenerate Fermi level occupancies we
  performed CPMC calculations. We used the full set
of spatial symmetries within the symmetrized version of PIRG (QP-PIRG),
which has been shown to significantly improve the
  results compared to earlier PIRG calculations \cite{Mizusaki04a}.
The symmetries we used for the 32 site lattice
were translations, a $\pi$ rotation, and a glide-plane symmetry.  We
also projected out the even spin parity state. The PIRG basis size was
up to $L$=512 Slater determinants for 32 sites and $L$=768
  for 64 sites.

\subsection{Spin structure factor}

\begin{figure}[tb]
  \centerline{\resizebox{3.3in}{!}{\includegraphics{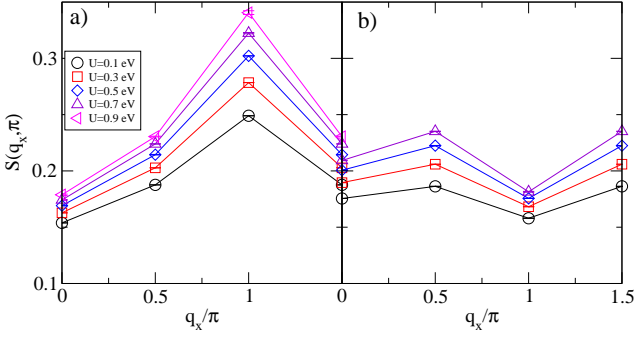}}}
  \caption{(color online) Dimer spin structure factor $S({\bf q})$ as
    a function of wavevector calculated using PIRG
    for $\rho$=1.5. Wavevectors are defined in
    terms of the effective dimer lattice with axes {\bf x} and {\bf y}
    as shown in Fig.~\ref{fig-lattice}. Panels are (a) 32 sites, $\kappa$-Cl,
    (b) 32 sites, $\kappa$-CN} 
  \label{fig-sfac}
\end{figure}
AFM is best explained within the effective dimer model \cite{Kanoda11a,Vojta99a,Schmalian98a,Kino98a,Kondo98a,Powell05a,Gan05a,Sahebsara06a,Watanabe06a,Kyung06a,Clay08a}, where the charge densities on
  the molecules of the dimer are equal.  Accordingly we
define the total $z$-component of spin on dimer $i$ as
\begin{equation}
  S_i^z  = \frac{1}{2} (n_{i_1,\uparrow} + n_{i_2,\uparrow}
  - n_{i_1,\downarrow} - n_{i_2,\downarrow}).
  \label{dimerspin}
\end{equation}  
In Eq.~\ref{dimerspin}, $i_1$ and $i_2$ refer to the two
different molecules within the dimer  $i$ and
$n_{j,\sigma}=c^\dagger_{j,\sigma}c_{j,\sigma}$. We  calculate the
dimer spin structure factor defined as
\begin{equation}
  S({\bf q})=\frac{1}{N_d} \sum_{j,k} e^{i{\bf q}\cdot({\bf r_j}-{\bf r_k})}
  \langle S^z_j S^z_k\rangle,
  \label{sfac}
\end{equation}
where $N_d$ is the number of dimers and dimer position vectors
{\bf r$_j$} 
are defined on a conventional square lattice, whose {\bf x} and
{\bf y} axes are indicated on Fig.~\ref{fig-lattice}.

In Fig.~\ref{fig-sfac} we show results for $S({\bf q})$ for both
$\kappa$-Cl (Fig.~\ref{fig-sfac}(a)) and $\kappa$-CN
(Fig.~\ref{fig-sfac}(b)).  $S({\bf q})$ for $\kappa$-Cl has a peak at
{\bf q}=$(\pi,\pi)$ consistent with N\'eel AFM correlations as
expected in the $\rho=1$ effective model for moderate frustration. As
shown in Fig.~\ref{fig-sfac}(a), the $(\pi,\pi)$ peak for $\kappa$-Cl
grows with increasing $U$. In contrast, we found no clear magnetic
ordering peak in $S({\bf q})$ for $\kappa$-CN, consistent with the
greater frustration within the effective dimer model in this case.

For the 64 site $\kappa$-Cl lattice $S(\pi,\pi)$ is smaller in magnitude
than for 32 sites, and the peak appears somewhat broader in momentum
space.  At present we do not have access to large enough system sizes
to perform a finite-size scaling for $S({\bf q})$, but the decrease of
$S(\pi,\pi)$ with increasing system size indicates that our
$\kappa$-Cl results are consistent with a metallic state with AFM {\it
  correlations} rather than an AFM insulating state ($S(\pi,\pi)/N_d$
should scale to a non-zero value in the presence of long-range AFM
order at $T=0$ in the thermodynamic limit). We discuss this issue
further in Section \ref{sec-discussion}. For the 64 site
  $\kappa$-CN lattice, we find that the variation $S({\bf q})$
  with {\bf q} is {\it less} than for 32 sites; this behavior is
  consistent with lack of magnetic order in the Mott insulating state
  of $\kappa$-CN.

\subsection{Pair-pair correlations}
\label{sec-pair}

\begin{figure}[htb!]
  \centerline{\resizebox{1.5in}{!}{\includegraphics{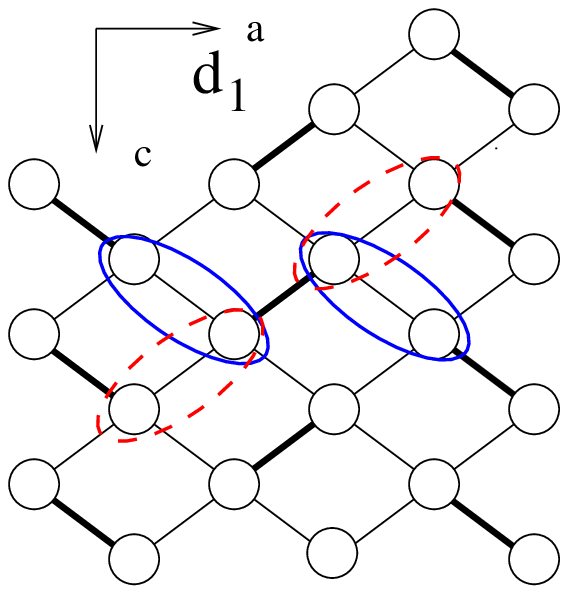}}\hspace{0.1in}%
    \resizebox{1.5in}{!}{\includegraphics{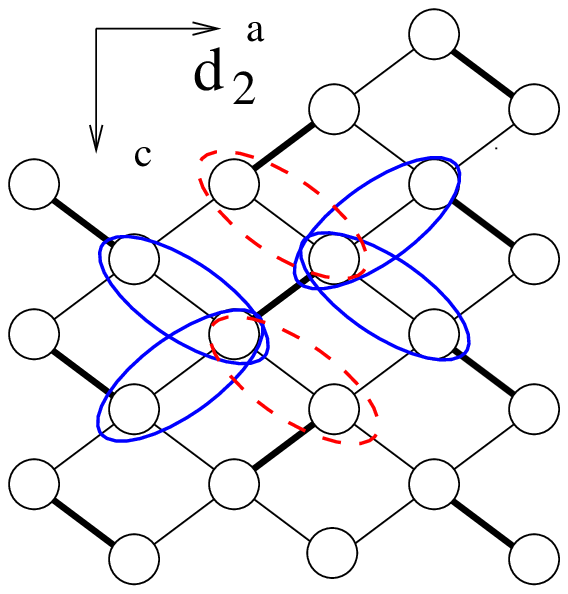}}}
  \centerline{\resizebox{1.5in}{!}{\includegraphics{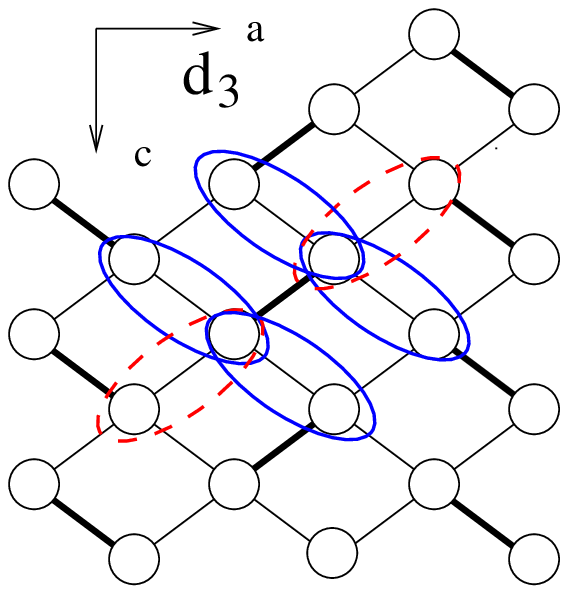}}\hspace{0.1in}%
    \resizebox{1.5in}{!}{\includegraphics{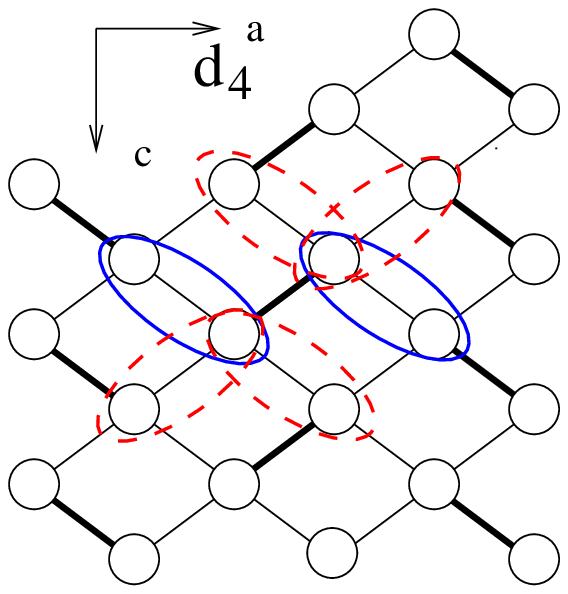}}\hspace{0.2in}}
  \caption{(color online) Pairing symmetries considered in our
    calculations. Each ellipse surrounding two sites indicates the
    location of a singlet in the superposition for the pair operator
    $\Delta^\dagger$. Blue, solid (red, dashed) singlets have opposite
    signs.}
    \label{symmetry}
\end{figure}

We calculate equal-time superconducting pair-pair correlations
$P_{ij}=\langle \Delta^\dagger_i \Delta_j \rangle$, where
$\Delta^\dagger_i$ is creates a superconducting pair centered at dimer
$i$.  There are two requirements for a complete theory of
correlated-electron superconductivity \cite{Gomes15a}: (i) e-e
correlations should enhance the value of $P_{ij}$ compared to its
uncorrelated value, and (ii) $P_{ij}$ has long-range order at $T=0$.
Here we focus on (i). 

\begin{figure}[tb!]
  \psfrag{label1}[B][B][2.5]{$\Theta_P$}
  \centerline{\resizebox{3.3in}{!}{\includegraphics{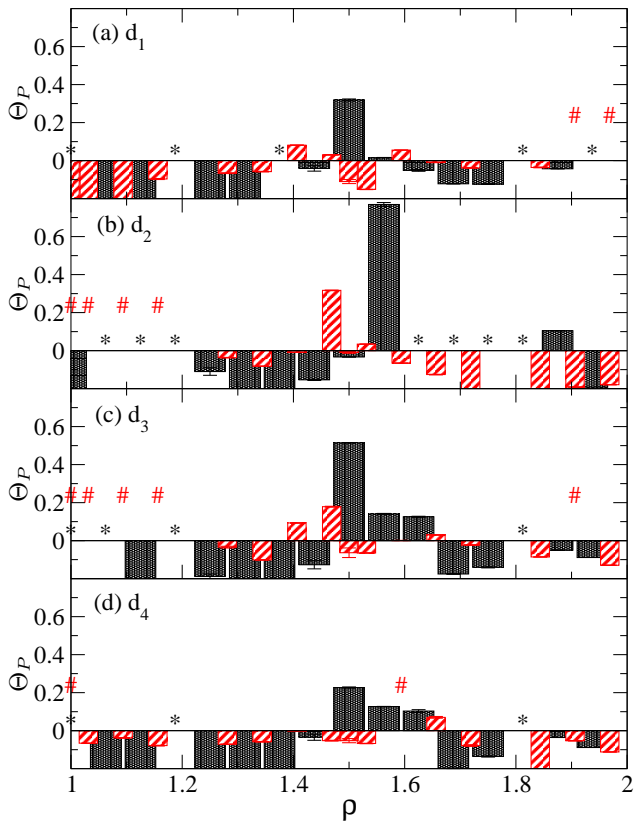}}}
  \caption{(color online) The enhancement factor $\Theta_P$ for the
    long-range component of the pair-pair correlation ($\Theta_P>0$
    implies pair-pair correlations enhanced over their $U=0$ values,
    see text), as a function of $\rho$, for the $\kappa$-Cl system for
    $U$=0.5 eV. Pair symmetries are (a) d$_1$, (b) d$_2$, (c) d$_3$,
    and (d) d$_4$ as defined in Fig.~\ref{symmetry}.  Shaded (striped)
    bars are for 32 (64) site lattices.  The symbols `*' and `\#'
    indicate densities not shown, for 32 and 64 sites, respectively;
    finite-size effects are particularly strong at these excluded
    $\rho$.  Pair-pair correlations are suppressed by $U$ at
these excluded $\rho$, precluding pairing; see 
Supplemental Material \cite{sm}.}
  \label{fig-pbar-cl}
\end{figure}
As mentioned above, previous works have shown (i) suppression of
pair-pair correlations within the effective $\rho=1$ model
\cite{Clay08a,Tocchio09a,Dayal12a,Gomes13a}, and (ii) the possibility
of a fluctuating CO 
within models focusing on the monomer molecules
\cite{Li10a,Dayal11a,Gomi10a,Naka10a,Hotta10a}.  We therefore
construct pair creation operators that allow unequal charge densities
on the monomer molecules that constitute a dimer in
Fig.~\ref{fig-lattice}.  We consider the central dimer (molecules
labeled 1 and 2) in Fig.~\ref{fig-lattice}. The pair operator
$\Delta^\dagger_1$ for this dimer is the superposition of singlets
between sites 1 and 2 and the surrounding sites 3$\cdots$8.  In order
to restrict the number of terms in $\Delta^\dagger_i$ and simplify the
calculation, we restrict the singlets to only the stronger 
interdimer bonds of
the lattice, {\it i.e.} the $t_{b_2}$ and $t_p$ bonds
(we ignore the weak $t_q$ bonds).
As an example, $d_{x^2-y^2}$ singlet pairs (labeled $d_1$ here, see below)
similar to the conventional
\begin{figure}[tb]
  \psfrag{label1}[B][B][2.5]{$\Theta_P$}
  \centerline{\resizebox{3.3in}{!}{\includegraphics{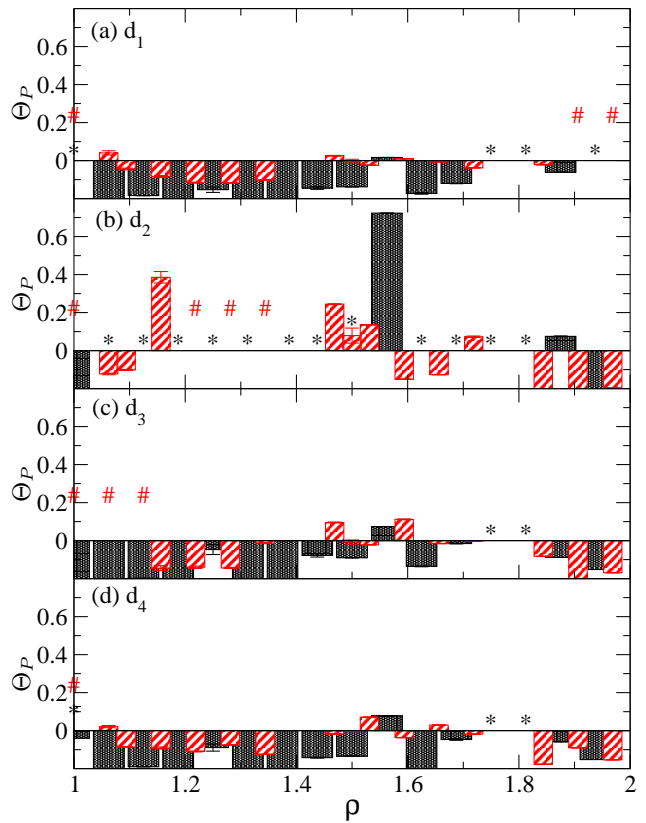}}}
  \caption{(color online) Same as in Fig.~\ref{fig-pbar-cl}, but for
    parameters for $\kappa$-CN. As in Fig.~\ref{fig-pbar-cl}, behavior of
all pair-pair correlations against $U$, including those for the excluded $\rho$
are shown in the Supplemental Material \cite{sm}. See text regarding
the peak in panel (b) at $\rho\approx 1.16$; we believe the apparent
  enhancement here is a finite-size effect.}
  \label{fig-pbar-cn}
\end{figure}
definition in the square effective lattice can be defined as follows
for the dimer (1,2) in Fig.~\ref{fig-lattice},
\begin{eqnarray}
  \Delta^\dagger_{d_1} &=&\frac{1}{2}\left[\frac{1}{\sqrt{2}}(c^\dagger_{1,\uparrow}
    c^\dagger_{8,\downarrow}-c^\dagger_{1,\downarrow}c^\dagger_{8,\uparrow})\right. \nonumber \\
    &-&  \frac{1}{\sqrt{2}}(c^\dagger_{1,\uparrow}
    c^\dagger_{3,\downarrow}-c^\dagger_{1,\downarrow}c^\dagger_{3,\uparrow}) \nonumber \\
    &+&  \frac{1}{\sqrt{2}}(c^\dagger_{2,\uparrow}
    c^\dagger_{5,\downarrow}-c^\dagger_{2,\downarrow}c^\dagger_{5,\uparrow}) \nonumber \\
    &-& \left.\frac{1}{\sqrt{2}}(c^\dagger_{2,\uparrow}
    c^\dagger_{6,\downarrow}-c^\dagger_{2,\downarrow}c^\dagger_{6,\uparrow})\right].
\label{pairing-eq}
\end{eqnarray}

 Given that the monomer lattice deviates strongly from the square
 lattice geometry several other pair symmetries are possible.
 Fig.~\ref{symmetry} summarizes the pair symmetries we considered.
These include four types of $d$-wave pairing (defined as symmetries with four nodes), with
four as well as six neighbors. We do not show the results for $s$-wave pairing symmetries,
as suppression of pair correlations were found with these, with  
four or six neighbors.
The difference between the four
 $d$-wave pair symmetries we consider is in the locations of the
 nodes, which we discuss further in Section
   \ref{sec-discussion}.

We calculate the average long-range value of the pair-pair
correlations\cite{Huang01a,Gomes15a} on each lattice,
$\bar{P}=1/N_P\sum_{|r_{ij}|>2} P_{ij}$. Here $N_p$ is the number of
terms in the sum, and distances are defined in units of the nearest
neighbor lattice distance of the effective dimer lattice.  The
restricted sum in the definition of $\bar{P}$ is necessary to
disentangle AFM and SC correlations. For the 32-site cluster for
example, there are five $P_{ij}$ separated by two or more lattice
spacings in the equivalent 4$\times$4 effective model.  

In order to compare the extents of enhancements of pair-pair
correlations by the Hubbard $U$ at different densities we normalize
$\bar{P}$ by its value for $U=0$, and show results for the enhancement
factor $\Theta_P=[\bar{P}(U)/\bar{P}(U=0)]-1$.
\begin{figure}[tb]
  \psfrag{label1}[B][B][3.0]{$\mathsf{\bar{P}}$}
  \centerline{\resizebox{3.3in}{!}{\includegraphics{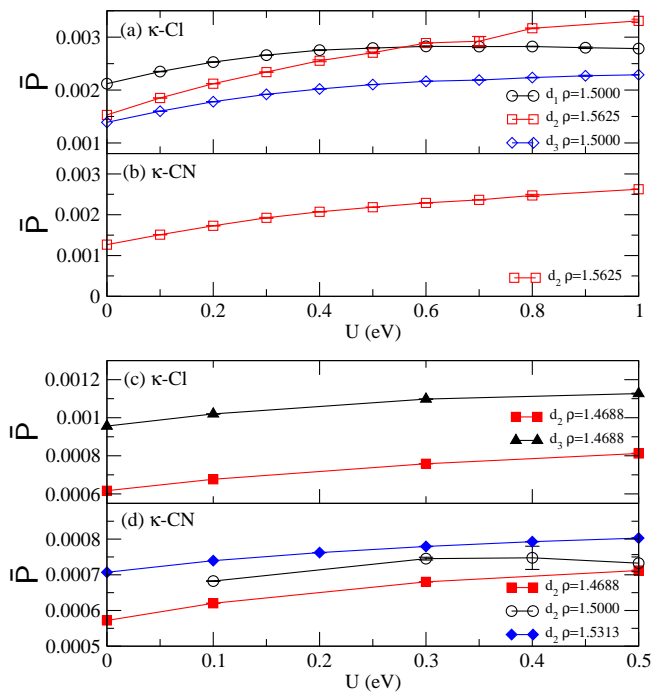}}}
  \caption{(color online) $U$ dependence of $\bar{P}$ for densities
    showing  significant enhancement of pair-pair correlations,
    for (a) $\kappa$-Cl, 32 sites, (b) $\kappa$-CN, 32 sites,
    (c) $\kappa$-Cl, 64 sites, and (d) $\kappa$-CN, 64 sites. all panels
  open (filled) symbols were calculated with PIRG (CPMC).}
    \label{fig-pbar-u}
\end{figure}
In Figs.~\ref{fig-pbar-cl} and ~\ref{fig-pbar-cn} we have shown
$\Theta_P$ as a function of $\rho$ for $U=0.5$ eV.  The normalization
of $\bar{P}$ fails for certain densities, where due to finite-size
effects $\bar{P}(U=0)$ is identically zero or very small in
magnitude. For this reason, in Fig.~\ref{fig-pbar-cl} and
Fig.~\ref{fig-pbar-cn},  we have excluded densities for which 
pair-pair correlations are very small in magnitude at nonzero $U$, or (b) pair-pair
correlations are negative.
The complete data including the points excluded in
Figs.~\ref{fig-pbar-cl} and ~\ref{fig-pbar-cn} for both the
$\kappa$-Cl and $\kappa$-CN lattices, for 32 as well as 64 sites are
shown in the Supplemental Material\cite{sm}.  As seen there suppression of
pair-pair correlations with $U$ occurs at any $\rho$ that has
been excluded.  Most of the data we excluded are also for densities
significantly away from $\rho=1.5$.

The results of Figs.~\ref{fig-pbar-cl} and \ref{fig-pbar-cn} are
remarkable, from multiple perspectives. First, in both cases
suppression of $\bar{P}$ is observed at all $\rho$ except at or near
$\rho=1.5$, where there occur enhancements of $\bar{P}$.  We are
ignoring the enhancement seen in the 64-site data for
Fig.~\ref{fig-pbar-cn}(b). The full $U$-dependence for this point is
in the Supplemental Material\cite{sm}, Fig.~S28. At this density, a
discontinuous transition occurs at small $U$, suggesting a
bandstructure effect. Furthermore at this $\rho$, $\bar{P}$ for the
$d_2$ symmetry is much smaller (but slightly above our cutoff) than
for other symmetries. This and the fact that we do not see enhancement
in any of our other results in the same density region suggests that
it is a finite-size effect.  Second, the strongest pairing enhancement
occurs for the $d_2$ symmetry for both 32 and 64 sites. Finally, only
for the $d_2$ pairing symmetry strong enhancement of $\bar{P}$ at
$\rho \simeq 1.5$ occurs for both the $\kappa$-Cl and $\kappa$-CN
lattice parameters.  This is a highly significant result, for as
remarked above, the $\kappa$-Cl and $\kappa$-CN have different $U=0$
single-particle level degeneracies at $\rho=1.5$. It gives us
confidence that the enhancement in pair correlations found here is not
an artifact of the one-electron band structure.

In Fig.~\ref{fig-pbar-u} we show the complete $U$-dependence of
$\bar{P}$ at the densities where significant enhancements in pair-pair
correlation occur, for both 32 and 64-site lattices, for both
$\kappa$-Cl and $\kappa$-CN structures. Compared to the 32 site data,
$\bar{P}$ as well as $\Theta_P$ are smaller in magnitude for 64 sites,
although we do expect that in the 64 site lattice
$\bar{P}$ will continue to increase at $\rho\approx 1.5$ for $U>0.5$
eV. The $\Theta_P$ data however indicate the absence of
true long-range superconducting order within our purely electronic
model.  If long-range superconducting order were present, $\bar{P}$
would have reached a constant value with increasing system size, while
$\bar{P}(U=0)$ decreased, in which case $\Theta_P$ would be expected
to increase with lattice size.  The enhancement of pair-pair
correlations uniquely at $\rho \simeq 1.5$ is nevertheless
significant, because {\it this is precisely the carrier concentration
  in the superconducting $\kappa$-(ET)$_2$X.} We elaborate on this
aspect of our result further in the following section.

\section{Discussions}
\label{sec-discussion}

We summarize our most significant results in this section, and discuss the 
implications of our work for
$\kappa$-(ET)$_2$X in particular, and for the family of 2D organic CTS
in general.

\subsection{AFM correlations versus long range AFM, and proximity to other broken symmetries}
\
The calculated decrease of $S(\pi,\pi)$ with increasing lattice size
for the $\kappa$-Cl lattice suggests that the ground state of the
present model does not have long-range AFM order, but is rather
metallic with short-range AFM correlations. We did not find any
evidence for a quantum phase transition to an AFM state in the 32-site
lattice up to $U\approx 1$ eV. While this conclusion appears
counterintuitive, given the strong emphasis on AFM in theoretical
works on these materials, it is in agreement with the experimental
behavior of the $\kappa$-(ET) family as a whole.  Experimentally,
$\kappa$-Cl and deuterated $\kappa$-Br are the only compounds that
exhibit AFM \cite{Kanoda11a}, and all other compounds are either
ambient pressure superconductors ($\kappa$-Br \cite{Ishiguro} and
$\kappa$-NCS \cite{Ishiguro}), QSL ($\kappa$-CN \cite{Kanoda11a}) or
PEC ($\kappa$-Hg(SCN)$_2$Cl \cite{Drichko14a}). Several of the more
complicated $\kappa$ materials such as
$\kappa$-(ET)$_4$[M(CN)$_6$][N(C$_2$H$_5$)$_4$]3H$_2$O
\cite{Magueres96a,Swietlik03a} and
$\kappa$-(ET)$_4$[M(CN)$_6$][N(C$_2$H$_5$)$_4$]2H$_2$O
\cite{Ota07a,Lapinski13a} (M = Co, Fe and Cr) are charge-ordered.
Charge-ordering in the last group of materials is accompanied by
spin-singlet formation and is apparently driven by strong interdimer
coupling \cite{Swietlik03a,Ota07a}, which is also the condition for
PEC formation \cite{Li10a,Dayal11a}.  Taken together, these results
suggest that even as AFM spin-spin correlations in $\kappa$-ET are
significant, these systems are at the threshold of transitions to
proximate broken symmetries that include the PEC as well as SC. This
observation is reminiscent of the occurrence of a CO phase competing
with both AFM and SC in the cuprates\cite{Chang12a,Wu11a,Blanco-Canosa14a,Hucker14a,Comin14a,SilvaNeto14a,SilvaNeto15a,Wu13a}, mentioned in Section \ref{sec-introduction}.

We speculate that the origin of long range AFM is due to either the
nearest neighbor Coulomb interaction $V$, or the coupling between
ET$_2^+$ cations and anions, both of which have been ignored in our
calculations. Because of the triangular dimer lattice, interdimer
Coulomb interactions are nearly the same in all directions and the
consequence of interdimer $V$ is small;  conversely, intradimer $V$
promotes single electron occupancy of dimers, and enhances
AFM. Similarly, it is conceivable that cation-anion coupling
determines the extent of electron localization in the cation layer.
The role of anions in the electrodynamics of $\kappa$-CN, for example,
has been emphasized by Dressel {\it et al.} \cite{Dressel16a}.

\subsection{Enhancement of pair-pair correlations and carrier density}

Our most significant result is the calculated enhancement of
pair-pair correlations by Hubbard $U$ within a narrow electron density
range about $\rho=1.5$.  Enhanced pair-pair correlations is a necessary
though not sufficient condition for SC.  
We previously used this criterion to evaluate the possibility of SC within the $\rho=1$
Hubbard model on triangular lattices
\cite{Clay08a,Dayal12a,Gomes13a}. Suppression of
pair-pair correlations by the Hubbard $U$ was found for all the
lattices we investigated. To the best of our knowledge, our results in Reference
\onlinecite{Gomes15a} for the first time showed an enhancement of
pair-pair correlations within the single-band Hubbard model in large
2D clusters (up to 100 sites).  It is then remarkable that we find here
enhanced superconducting pair-pair correlations for two different
$\kappa$ lattices ($\kappa$-Cl and $\kappa$-CN), for two different lattice sizes in each case,
{\it for precisely the same narrow carrier concentration range that would
be anticipated from Reference \onlinecite{Gomes15a}.} This is
particularly so considering the relevance of this carrier density to
experimental (ET)$_2$X.

Within our theory, AFM is a signature of
strong correlations, but is not the driver of electron
pairing. Enhanced pair-pair correlations originate from the strong
tendency to spin-singlet coupling at $\rho=0.5$ and 1.5, both because
of the existence of a commensurate PEC at these densities
\cite{Li10a,Dayal11a}, and because the stabilization due to the kinetic
energy gain from pair motion is highest at these carrier
concentrations. Consideration of nearest neighbor spin-bonded sites as effective ``negative
U'' centers has a long history, especially in the context of bipolaron theories of SC \cite{Micnas90a}. 
Such a ``mapping'' for $\rho$ exactly 0.5 or 1.5 leads to an effective $\frac{1}{2}$-filled band 
of bosons \cite{Mazumdar08a}, with nearly degenerate CO and SC. 
We emphasize
that the effective ``negative U'' model is different from the existing
effective $\frac{1}{2}$-filled band theories emphasizing AFM
\cite{Vojta99a,Schmalian98a,Kino98a,Kondo98a,Powell05a,Gan05a,Sahebsara06a,Watanabe06a,Kyung06a,Hebert15a},
within which no CO phase is anticipated.
Thus within our theory, SC proximate to AFM
(as occurs in the $\kappa$-(ET)$_2$X), as well as to CO (as occurs in
other crystal structures or in the anionic superconductors
\cite{Andres05a,Bangura05a,Tamura06a,Shikama12a}) are {\it
  manifestations of the same correlation effects and to be
  anticipated}. Of course, the CO in these cases should have the
charge pattern of the PEC (as opposed to the Wigner crystal) with
ground state spin singlet character.  

\subsection{PEL versus SC}

Our work indicates that while repulsive e-e interactions can drive the
transition to a PEL with short range pair-pair correlations at $\rho
\simeq 1.5$, additional interactions missing in the purely electronic
Hubbard model will be necessary to obtain long-range superconducting
correlations. The most likely such interactions are that between the
electrons and lattice vibrations involving intramolecular Holstein
phonons \cite{Holstein59a} and intermolecular Su-Schrieffer-Heeger
(SSH) phonons \cite{Su79a}. We emphasize that there are
many counterexamples to the notions that e-e and electron-phonon (e-p)
interactions invariably negate each others effects and
that e-p coupling can only generate SC of $s$-wave symmetry. One
widely known counterexample is the enhancement of the e-p interaction
driven Peierls bond-alternation in the one-dimensional half-filled
band by e-e interactions \cite{Mazumdar83b,Hirsch83b,Mazumdar85a}.
With the Holstein-type e-p coupling, integrating out the
phonon degrees of freedom leads to an effective negative $U$
interaction giving onsite CDW order or $s$-wave SC \cite{Hirsch83a}.  However,
for the 2D dispersionless SSH model, integrating out the phonons leads
to an attractive term proportional to the square of the kinetic
energy \cite{Assaad96a}.  This term mediates $d_{x^2-y^2}$ SC and
competes with AFM mediated by $U$ at $\rho=1$
\cite{Assaad96a,Assaad97a,Assaad98a}.

We have similarly found co-operative interactions between the effects
of e-e and e-p interactions in the formation of the PEC in both 1D
\cite{Clay03a} and 2D \cite{Li10a,Dayal11a}. In all these cases the
retarded phonon interactions can be thought as ``following'' the
instantaneous correlations driven by the e-e interactions. Thus the
interpretation of our result that the PEL is unique to $\rho \simeq
1.5$ should be that in the presence of e-p interactions this is the
carrier density in the $\kappa$ lattice where correlated-electron SC
is most likely.

The idea that AFM coupling may be driving the formation of nearest
neighbor Cooper pairs, whose mobility is then enhanced by e-p
interactions has also been proposed within later versions of the
bipolaron theory of SC \cite{Hague07a,Vidmar10a}. These calculations
are for only two electrons on a lattice. The significant advancement
reached in our work is that we have demonstrated here the formation of
similar mobile bipolarons within the electron-only Hamiltonian for the
full many-electron system as opposed to one with only two electrons.

\subsection{Symmetry of superconducting order parameter}

Experiments using a wide range of probes suggest that the SC pairing
throughout the $\kappa$-(ET)$_2$X family is singlet with nodes in the
order parameter. Site-selective $^{13}$C NMR experiments have been
performed on the ambient-pressure superconductors $\kappa$-Br
\cite{deSoto95a,Kanoda96a} $\kappa$-NCS \cite{Miyagawa04a}, and also
for $\kappa$-CN under pressure in its superconducting state
\cite{Shimizu10a}.  Common features found from the NMR experiments
include singlet pairing, no coherence peak in 1/$T_1$, and 1/$T_1
\propto T^3$ at low temperatures. These suggest a non-BCS mechanism
and the presence of nodes. In contrast to early specific heat
measurements that supported $s$-wave pairing
\cite{Elsinger00a,Muller02b}, more recent measurements are consistent
with a nodal order parameter \cite{Taylor07a}. Microwave penetration
depth measurements \cite{Milbradt13a,Perunov12a} as well as STM
tunneling experiments
\cite{Arai00a,Arai01a,Ichimura03a,Ichimura05a,Ichimura08a,Oka15a} are
also consistent with nodes.

While experiments are generally in agreement that the SC is singlet
and has a nodal order parameter, there is less agreement on the
specific form of the order parameter and the location of nodes in the
conducting plane.  Candidate symmetries  $d_{x^2-y^2}$ and $d_{xy}$
 differ in the locations of their nodes; in the experimental
literature $d_{x^2-y^2}$ symmetry is usually assumed to have nodes at
45$^\circ$ to the crystal axes, while $d_{xy}$ has nodes along the
crystal axes. It is important to note that in theoretical work based
on the effective half-filled model, these two symmetries are
interchanged, as the effective $x$ and $y$ axes are {\it rotated} with
respect to the crystal axes (see Fig.~\ref{fig-lattice}).
Magneto-optical \cite{Schrama99a} and specific heat measurements in a
magnetic field find the nodes to coincide with the crystal axes
\cite{Malone10a}.  Thermal conductivity \cite{Izawa01a} and STM
measurements \cite{Arai01a,Ichimura08a} however find the nodes between
the crystal axes, although STM measurements on a partially deuterated
$\kappa$-Br suggest a mixture of two order parameters \cite{Oka15a}.
Experiments sensitive to the position of the nodes have not been
performed on $\kappa$-CN, which is superconducting only under pressure. 

In our calculations (see Fig.~\ref{symmetry}), the $d_1$ symmetry has
nodes along the crystal axes, while the $d_2$ symmetry instead has
nodes at an angle between the crystal axes. The $d_3$ and $d_4$ have
one node along a crystal axis and one off-axis.  In our results we
found the strongest enhancement for the symmetry $d_2$, which is also
the only symmetry with enhanced pairing for both $\kappa$-Cl and
$\kappa$-CN.  However, Fig.~\ref{fig-pbar-cl} shows that several
pairing symmetries are enhanced for $\kappa$-Cl, suggesting the
possibility that the optimum pairing symmetry may vary  for
different X in the $\kappa$-ET series.

\section{Conclusions}
\label{sec-conclusion}

To summarize, from numerical calculations on the Hubbard model on the
monomer lattice of $\kappa$-(ET)$_2$X solids we have found a PEL state
with enhanced superconducting pair-pair correlations exactly for the
cationic charge where SC is found experimentally. We have also demonstrated
that the pair-pair correlations are suppressed by the Hubbard
interaction at all other carrier densities. The superconducting order
we find is short-range, and considerable work involving both
e-e and e-p interactions will be necessary
before a complete theory of SC in the CTS is reached. To the best of
our knowledge, however, robust Coulomb enhancement of pair
correlations has not been found before. Taken together with our
previous work \cite{Gomes15a}, this gives us confidence that the
physical ideas behind this work, viz., (i) enhancement of
superconducting pair correlations requires a proximate spin-singlet
state in the phase diagram, and (ii) such a spin-singlet is strongly stabilized
in 2D for $\rho=0.5$ or 1.5, are fundamentally correct. The spin-singlet
state in our case has a strong tendency to form a CO. As pointed out
earlier, CO proximate to SC has been found within the pseudogap phase
of the hole-doped cuprates. We are currently investigating the
implication of our results for the cuprates.

\section{Acknowledgments}
\label{Ack}

S. M. and R. T. C. acknowledge very useful discussions with M. Dressel
and S. Tomic, and for sending them preprints of references
\onlinecite{Sedlmeier12a} and \onlinecite{Dressel16a}.  They are also
grateful to N. Drichko for sending them the preprint of reference
\onlinecite{Drichko14a}.  This work was supported by the US Department
of Energy grant DE-FG02-06ER46315. N. G. was supported by NSF grant
CHE-1151475.  Part of the calculations were performed using resources
of the National Energy Research Scientific Computing Center (NERSC),
which is supported by the Office of Science of the U.S. Department of
Energy under Contract No. DE-AC02-05CH11231.


\begin{thebibliography}{101}%
\makeatletter
\providecommand \@ifxundefined [1]{%
 \@ifx{#1\undefined}
}%
\providecommand \@ifnum [1]{%
 \ifnum #1\expandafter \@firstoftwo
 \else \expandafter \@secondoftwo
 \fi
}%
\providecommand \@ifx [1]{%
 \ifx #1\expandafter \@firstoftwo
 \else \expandafter \@secondoftwo
 \fi
}%
\providecommand \natexlab [1]{#1}%
\providecommand \enquote  [1]{``#1''}%
\providecommand \bibnamefont  [1]{#1}%
\providecommand \bibfnamefont [1]{#1}%
\providecommand \citenamefont [1]{#1}%
\providecommand \href@noop [0]{\@secondoftwo}%
\providecommand \href [0]{\begingroup \@sanitize@url \@href}%
\providecommand \@href[1]{\@@startlink{#1}\@@href}%
\providecommand \@@href[1]{\endgroup#1\@@endlink}%
\providecommand \@sanitize@url [0]{\catcode `\\12\catcode `\$12\catcode
  `\&12\catcode `\#12\catcode `\^12\catcode `\_12\catcode `\%12\relax}%
\providecommand \@@startlink[1]{}%
\providecommand \@@endlink[0]{}%
\providecommand \url  [0]{\begingroup\@sanitize@url \@url }%
\providecommand \@url [1]{\endgroup\@href {#1}{\urlprefix }}%
\providecommand \urlprefix  [0]{URL }%
\providecommand \Eprint [0]{\href }%
\providecommand \doibase [0]{http://dx.doi.org/}%
\providecommand \selectlanguage [0]{\@gobble}%
\providecommand \bibinfo  [0]{\@secondoftwo}%
\providecommand \bibfield  [0]{\@secondoftwo}%
\providecommand \translation [1]{[#1]}%
\providecommand \BibitemOpen [0]{}%
\providecommand \bibitemStop [0]{}%
\providecommand \bibitemNoStop [0]{.\EOS\space}%
\providecommand \EOS [0]{\spacefactor3000\relax}%
\providecommand \BibitemShut  [1]{\csname bibitem#1\endcsname}%
\let\auto@bib@innerbib\@empty
\bibitem [{\citenamefont {Kanoda}\ and\ \citenamefont
  {Kato}(2011)}]{Kanoda11a}%
  \BibitemOpen
  \bibfield  {author} {\bibinfo {author} {\bibfnamefont {K.}~\bibnamefont
  {Kanoda}}\ and\ \bibinfo {author} {\bibfnamefont {R.}~\bibnamefont {Kato}},\
  }\href@noop {} {\bibfield  {journal} {\bibinfo  {journal} {Annu. Rev.
  Condens. Matter Phys.}\ }\textbf {\bibinfo {volume} {2}},\ \bibinfo {pages}
  {167} (\bibinfo {year} {2011})}\BibitemShut {NoStop}%
\bibitem [{\citenamefont {Ishiguro}\ \emph {et~al.}(1998)\citenamefont
  {Ishiguro}, \citenamefont {Yamaji},\ and\ \citenamefont {Saito}}]{Ishiguro}%
  \BibitemOpen
  \bibfield  {author} {\bibinfo {author} {\bibfnamefont {T.}~\bibnamefont
  {Ishiguro}}, \bibinfo {author} {\bibfnamefont {K.}~\bibnamefont {Yamaji}}, \
  and\ \bibinfo {author} {\bibfnamefont {G.}~\bibnamefont {Saito}},\
  }\href@noop {} {\emph {\bibinfo {title} {Organic Superconductors}}}\
  (\bibinfo  {publisher} {Springer-Verlag},\ \bibinfo {address} {New York},\
  \bibinfo {year} {1998})\BibitemShut {NoStop}%
\bibitem [{\citenamefont {Kino}\ and\ \citenamefont
  {Fukuyama}(1995)}]{Kino95a}%
  \BibitemOpen
  \bibfield  {author} {\bibinfo {author} {\bibfnamefont {H.}~\bibnamefont
  {Kino}}\ and\ \bibinfo {author} {\bibfnamefont {H.}~\bibnamefont
  {Fukuyama}},\ }\href@noop {} {\bibfield  {journal} {\bibinfo  {journal} {J.\
  Phys.\ Soc.\ Jpn.}\ }\textbf {\bibinfo {volume} {64}},\ \bibinfo {pages}
  {2726} (\bibinfo {year} {1995})}\BibitemShut {NoStop}%
\bibitem [{\citenamefont {Anderson}(1987)}]{Anderson87a}%
  \BibitemOpen
  \bibfield  {author} {\bibinfo {author} {\bibfnamefont {P.~W.}\ \bibnamefont
  {Anderson}},\ }\href@noop {} {\bibfield  {journal} {\bibinfo  {journal}
  {Science}\ }\textbf {\bibinfo {volume} {235}},\ \bibinfo {pages} {1196}
  (\bibinfo {year} {1987})}\BibitemShut {NoStop}%
\bibitem [{\citenamefont {Scalapino}(1995)}]{Scalapino95a}%
  \BibitemOpen
  \bibfield  {author} {\bibinfo {author} {\bibfnamefont {D.~J.}\ \bibnamefont
  {Scalapino}},\ }\href@noop {} {\bibfield  {journal} {\bibinfo  {journal}
  {Phys. Rep.}\ }\textbf {\bibinfo {volume} {250}},\ \bibinfo {pages} {329}
  (\bibinfo {year} {1995})}\BibitemShut {NoStop}%
\bibitem [{\citenamefont {Anderson}\ \emph {et~al.}(2004)\citenamefont
  {Anderson}, \citenamefont {Lee}, \citenamefont {Randeria}, \citenamefont
  {Rice}, \citenamefont {Trivedi},\ and\ \citenamefont {Zhang}}]{Anderson04a}%
  \BibitemOpen
  \bibfield  {author} {\bibinfo {author} {\bibfnamefont {P.~W.}\ \bibnamefont
  {Anderson}}, \bibinfo {author} {\bibfnamefont {P.~A.}\ \bibnamefont {Lee}},
  \bibinfo {author} {\bibfnamefont {M.}~\bibnamefont {Randeria}}, \bibinfo
  {author} {\bibfnamefont {T.~M.}\ \bibnamefont {Rice}}, \bibinfo {author}
  {\bibfnamefont {N.}~\bibnamefont {Trivedi}}, \ and\ \bibinfo {author}
  {\bibfnamefont {F.~C.}\ \bibnamefont {Zhang}},\ }\href@noop {} {\bibfield
  {journal} {\bibinfo  {journal} {J. Phys. Condens. Matter}\ }\textbf {\bibinfo
  {volume} {16}},\ \bibinfo {pages} {R755} (\bibinfo {year}
  {2004})}\BibitemShut {NoStop}%
\bibitem [{\citenamefont {Vojta}\ and\ \citenamefont
  {Dagotto}(1999)}]{Vojta99a}%
  \BibitemOpen
  \bibfield  {author} {\bibinfo {author} {\bibfnamefont {M.}~\bibnamefont
  {Vojta}}\ and\ \bibinfo {author} {\bibfnamefont {E.}~\bibnamefont
  {Dagotto}},\ }\href@noop {} {\bibfield  {journal} {\bibinfo  {journal}
  {Phys.\ Rev.\ B}\ }\textbf {\bibinfo {volume} {59}},\ \bibinfo {pages} {R713}
  (\bibinfo {year} {1999})}\BibitemShut {NoStop}%
\bibitem [{\citenamefont {Schmalian}(1998)}]{Schmalian98a}%
  \BibitemOpen
  \bibfield  {author} {\bibinfo {author} {\bibfnamefont {J.}~\bibnamefont
  {Schmalian}},\ }\href@noop {} {\bibfield  {journal} {\bibinfo  {journal}
  {Phys.\ Rev.\ Lett.}\ }\textbf {\bibinfo {volume} {81}},\ \bibinfo {pages}
  {4232} (\bibinfo {year} {1998})}\BibitemShut {NoStop}%
\bibitem [{\citenamefont {Kino}\ and\ \citenamefont {Kontani}(1998)}]{Kino98a}%
  \BibitemOpen
  \bibfield  {author} {\bibinfo {author} {\bibfnamefont {H.}~\bibnamefont
  {Kino}}\ and\ \bibinfo {author} {\bibfnamefont {H.}~\bibnamefont {Kontani}},\
  }\href@noop {} {\bibfield  {journal} {\bibinfo  {journal} {J.\ Phys.\ Soc.\
  Jpn.}\ }\textbf {\bibinfo {volume} {67}},\ \bibinfo {pages} {3691} (\bibinfo
  {year} {1998})}\BibitemShut {NoStop}%
\bibitem [{\citenamefont {Kondo}\ and\ \citenamefont
  {Moriya}(1998)}]{Kondo98a}%
  \BibitemOpen
  \bibfield  {author} {\bibinfo {author} {\bibfnamefont {H.}~\bibnamefont
  {Kondo}}\ and\ \bibinfo {author} {\bibfnamefont {T.}~\bibnamefont {Moriya}},\
  }\href@noop {} {\bibfield  {journal} {\bibinfo  {journal} {J.\ Phys.\ Soc.\
  Jpn.}\ }\textbf {\bibinfo {volume} {67}},\ \bibinfo {pages} {3695} (\bibinfo
  {year} {1998})}\BibitemShut {NoStop}%
\bibitem [{\citenamefont {Powell}\ and\ \citenamefont
  {McKenzie}(2005)}]{Powell05a}%
  \BibitemOpen
  \bibfield  {author} {\bibinfo {author} {\bibfnamefont {B.~J.}\ \bibnamefont
  {Powell}}\ and\ \bibinfo {author} {\bibfnamefont {R.~H.}\ \bibnamefont
  {McKenzie}},\ }\href@noop {} {\bibfield  {journal} {\bibinfo  {journal}
  {Phys.\ Rev.\ Lett.}\ }\textbf {\bibinfo {volume} {94}},\ \bibinfo {pages}
  {047004} (\bibinfo {year} {2005})}\BibitemShut {NoStop}%
\bibitem [{\citenamefont {Gan}\ \emph {et~al.}(2005)\citenamefont {Gan},
  \citenamefont {Chen}, \citenamefont {Su},\ and\ \citenamefont
  {Zhang}}]{Gan05a}%
  \BibitemOpen
  \bibfield  {author} {\bibinfo {author} {\bibfnamefont {J.~Y.}\ \bibnamefont
  {Gan}}, \bibinfo {author} {\bibfnamefont {Y.}~\bibnamefont {Chen}}, \bibinfo
  {author} {\bibfnamefont {Z.~B.}\ \bibnamefont {Su}}, \ and\ \bibinfo {author}
  {\bibfnamefont {F.~C.}\ \bibnamefont {Zhang}},\ }\href@noop {} {\bibfield
  {journal} {\bibinfo  {journal} {Phys.\ Rev.\ Lett.}\ }\textbf {\bibinfo
  {volume} {94}},\ \bibinfo {pages} {067005} (\bibinfo {year}
  {2005})}\BibitemShut {NoStop}%
\bibitem [{\citenamefont {Sahebsara}\ and\ \citenamefont
  {\protect{S\'en\'echal}}(2006)}]{Sahebsara06a}%
  \BibitemOpen
  \bibfield  {author} {\bibinfo {author} {\bibfnamefont {P.}~\bibnamefont
  {Sahebsara}}\ and\ \bibinfo {author} {\bibfnamefont {D.}~\bibnamefont
  {\protect{S\'en\'echal}}},\ }\href@noop {} {\bibfield  {journal} {\bibinfo
  {journal} {Phys.\ Rev.\ Lett.}\ }\textbf {\bibinfo {volume} {97}},\ \bibinfo
  {pages} {257004} (\bibinfo {year} {2006})}\BibitemShut {NoStop}%
\bibitem [{\citenamefont {Watanabe}\ \emph {et~al.}(2006)\citenamefont
  {Watanabe}, \citenamefont {Yokoyama}, \citenamefont {Tanaka},\ and\
  \citenamefont {Inoue}}]{Watanabe06a}%
  \BibitemOpen
  \bibfield  {author} {\bibinfo {author} {\bibfnamefont {T.}~\bibnamefont
  {Watanabe}}, \bibinfo {author} {\bibfnamefont {H.}~\bibnamefont {Yokoyama}},
  \bibinfo {author} {\bibfnamefont {Y.}~\bibnamefont {Tanaka}}, \ and\ \bibinfo
  {author} {\bibfnamefont {J.}~\bibnamefont {Inoue}},\ }\href@noop {}
  {\bibfield  {journal} {\bibinfo  {journal} {J.\ Phys.\ Soc.\ Jpn.}\ }\textbf
  {\bibinfo {volume} {75}},\ \bibinfo {pages} {074707} (\bibinfo {year}
  {2006})}\BibitemShut {NoStop}%
\bibitem [{\citenamefont {Kyung}\ and\ \citenamefont
  {Tremblay}(2006)}]{Kyung06a}%
  \BibitemOpen
  \bibfield  {author} {\bibinfo {author} {\bibfnamefont {B.}~\bibnamefont
  {Kyung}}\ and\ \bibinfo {author} {\bibfnamefont {A.~M.~S.}\ \bibnamefont
  {Tremblay}},\ }\href@noop {} {\bibfield  {journal} {\bibinfo  {journal}
  {Phys.\ Rev.\ Lett.}\ }\textbf {\bibinfo {volume} {97}},\ \bibinfo {pages}
  {046402} (\bibinfo {year} {2006})}\BibitemShut {NoStop}%
\bibitem [{\citenamefont {Hebert}\ \emph {et~al.}(2015)\citenamefont {Hebert},
  \citenamefont {Semon},\ and\ \citenamefont {Tremblay}}]{Hebert15a}%
  \BibitemOpen
  \bibfield  {author} {\bibinfo {author} {\bibfnamefont {C.~D.}\ \bibnamefont
  {Hebert}}, \bibinfo {author} {\bibfnamefont {P.}~\bibnamefont {Semon}}, \
  and\ \bibinfo {author} {\bibfnamefont {A.~M.~S.}\ \bibnamefont {Tremblay}},\
  }\href@noop {} {\bibfield  {journal} {\bibinfo  {journal} {Phys.\ Rev.\ B}\
  }\textbf {\bibinfo {volume} {92}},\ \bibinfo {pages} {195112} (\bibinfo
  {year} {2015})}\BibitemShut {NoStop}%
\bibitem [{\citenamefont {Clay}\ \emph {et~al.}(2008)\citenamefont {Clay},
  \citenamefont {Li},\ and\ \citenamefont {Mazumdar}}]{Clay08a}%
  \BibitemOpen
  \bibfield  {author} {\bibinfo {author} {\bibfnamefont {R.~T.}\ \bibnamefont
  {Clay}}, \bibinfo {author} {\bibfnamefont {H.}~\bibnamefont {Li}}, \ and\
  \bibinfo {author} {\bibfnamefont {S.}~\bibnamefont {Mazumdar}},\ }\href@noop
  {} {\bibfield  {journal} {\bibinfo  {journal} {Phys.\ Rev.\ Lett.}\ }\textbf
  {\bibinfo {volume} {101}},\ \bibinfo {pages} {166403} (\bibinfo {year}
  {2008})}\BibitemShut {NoStop}%
\bibitem [{\citenamefont {Tocchio}\ \emph {et~al.}(2009)\citenamefont
  {Tocchio}, \citenamefont {Parola}, \citenamefont {Gros},\ and\ \citenamefont
  {Becca}}]{Tocchio09a}%
  \BibitemOpen
  \bibfield  {author} {\bibinfo {author} {\bibfnamefont {L.~F.}\ \bibnamefont
  {Tocchio}}, \bibinfo {author} {\bibfnamefont {A.}~\bibnamefont {Parola}},
  \bibinfo {author} {\bibfnamefont {C.}~\bibnamefont {Gros}}, \ and\ \bibinfo
  {author} {\bibfnamefont {F.}~\bibnamefont {Becca}},\ }\href@noop {}
  {\bibfield  {journal} {\bibinfo  {journal} {Phys.\ Rev.\ B}\ }\textbf
  {\bibinfo {volume} {80}},\ \bibinfo {pages} {064419} (\bibinfo {year}
  {2009})}\BibitemShut {NoStop}%
\bibitem [{\citenamefont {Dayal}\ \emph {et~al.}(2012)\citenamefont {Dayal},
  \citenamefont {Clay},\ and\ \citenamefont {Mazumdar}}]{Dayal12a}%
  \BibitemOpen
  \bibfield  {author} {\bibinfo {author} {\bibfnamefont {S.}~\bibnamefont
  {Dayal}}, \bibinfo {author} {\bibfnamefont {R.~T.}\ \bibnamefont {Clay}}, \
  and\ \bibinfo {author} {\bibfnamefont {S.}~\bibnamefont {Mazumdar}},\
  }\href@noop {} {\bibfield  {journal} {\bibinfo  {journal} {Phys.\ Rev.\ B}\
  }\textbf {\bibinfo {volume} {85}},\ \bibinfo {pages} {165141} (\bibinfo
  {year} {2012})}\BibitemShut {NoStop}%
\bibitem [{\citenamefont {Gomes}\ \emph {et~al.}(2013)\citenamefont {Gomes},
  \citenamefont {Clay},\ and\ \citenamefont {Mazumdar}}]{Gomes13a}%
  \BibitemOpen
  \bibfield  {author} {\bibinfo {author} {\bibfnamefont {N.}~\bibnamefont
  {Gomes}}, \bibinfo {author} {\bibfnamefont {R.~T.}\ \bibnamefont {Clay}}, \
  and\ \bibinfo {author} {\bibfnamefont {S.}~\bibnamefont {Mazumdar}},\
  }\href@noop {} {\bibfield  {journal} {\bibinfo  {journal} {J. Phys. Condens.
  Matter}\ }\textbf {\bibinfo {volume} {25}},\ \bibinfo {pages} {385603}
  (\bibinfo {year} {2013})}\BibitemShut {NoStop}%
\bibitem [{\citenamefont {Chang}\ \emph {et~al.}(2012)\citenamefont {Chang}
  \emph {et~al.}}]{Chang12a}%
  \BibitemOpen
  \bibfield  {author} {\bibinfo {author} {\bibfnamefont {J.}~\bibnamefont
  {Chang}} \emph {et~al.},\ }\href@noop {} {\bibfield  {journal} {\bibinfo
  {journal} {Nature Physics}\ }\textbf {\bibinfo {volume} {8}},\ \bibinfo
  {pages} {871} (\bibinfo {year} {2012})}\BibitemShut {NoStop}%
\bibitem [{\citenamefont {Wu}\ \emph {et~al.}(2011)\citenamefont {Wu},
  \citenamefont {Mayaffre}, \citenamefont {\protect{Kr\"amer}}, \citenamefont
  {\protect{Horvati\'c}}, \citenamefont {Berthier}, \citenamefont {Hardy},
  \citenamefont {Liang}, \citenamefont {Bonn},\ and\ \citenamefont
  {Julien}}]{Wu11a}%
  \BibitemOpen
  \bibfield  {author} {\bibinfo {author} {\bibfnamefont {T.}~\bibnamefont
  {Wu}}, \bibinfo {author} {\bibfnamefont {H.}~\bibnamefont {Mayaffre}},
  \bibinfo {author} {\bibfnamefont {S.}~\bibnamefont {\protect{Kr\"amer}}},
  \bibinfo {author} {\bibfnamefont {M.}~\bibnamefont {\protect{Horvati\'c}}},
  \bibinfo {author} {\bibfnamefont {C.}~\bibnamefont {Berthier}}, \bibinfo
  {author} {\bibfnamefont {W.~N.}\ \bibnamefont {Hardy}}, \bibinfo {author}
  {\bibfnamefont {R.}~\bibnamefont {Liang}}, \bibinfo {author} {\bibfnamefont
  {D.~A.}\ \bibnamefont {Bonn}}, \ and\ \bibinfo {author} {\bibfnamefont
  {M.-H.}\ \bibnamefont {Julien}},\ }\href@noop {} {\bibfield  {journal}
  {\bibinfo  {journal} {Nature}\ }\textbf {\bibinfo {volume} {477}},\ \bibinfo
  {pages} {191} (\bibinfo {year} {2011})}\BibitemShut {NoStop}%
\bibitem [{\citenamefont {Blanco-Canosa}\ \emph {et~al.}(2014)\citenamefont
  {Blanco-Canosa} \emph {et~al.}}]{Blanco-Canosa14a}%
  \BibitemOpen
  \bibfield  {author} {\bibinfo {author} {\bibfnamefont {S.}~\bibnamefont
  {Blanco-Canosa}} \emph {et~al.},\ }\href@noop {} {\bibfield  {journal}
  {\bibinfo  {journal} {Phys.\ Rev.\ B}\ }\textbf {\bibinfo {volume} {90}},\
  \bibinfo {pages} {054513} (\bibinfo {year} {2014})}\BibitemShut {NoStop}%
\bibitem [{\citenamefont {H\protect{\"u}cker}\ \emph
  {et~al.}(2014)\citenamefont {H\protect{\"u}cker} \emph {et~al.}}]{Hucker14a}%
  \BibitemOpen
  \bibfield  {author} {\bibinfo {author} {\bibfnamefont {M.}~\bibnamefont
  {H\protect{\"u}cker}} \emph {et~al.},\ }\href@noop {} {\bibfield  {journal}
  {\bibinfo  {journal} {Phys.\ Rev.\ B}\ }\textbf {\bibinfo {volume} {90}},\
  \bibinfo {pages} {054514} (\bibinfo {year} {2014})}\BibitemShut {NoStop}%
\bibitem [{\citenamefont {Comin}\ \emph {et~al.}(2014)\citenamefont {Comin}
  \emph {et~al.}}]{Comin14a}%
  \BibitemOpen
  \bibfield  {author} {\bibinfo {author} {\bibfnamefont {R.}~\bibnamefont
  {Comin}} \emph {et~al.},\ }\href@noop {} {\bibfield  {journal} {\bibinfo
  {journal} {Science}\ }\textbf {\bibinfo {volume} {343}},\ \bibinfo {pages}
  {390} (\bibinfo {year} {2014})}\BibitemShut {NoStop}%
\bibitem [{\citenamefont {Neto}\ \emph {et~al.}(2014)\citenamefont {Neto} \emph
  {et~al.}}]{SilvaNeto14a}%
  \BibitemOpen
  \bibfield  {author} {\bibinfo {author} {\bibfnamefont {E.~H. D.~S.}\
  \bibnamefont {Neto}} \emph {et~al.},\ }\href@noop {} {\bibfield  {journal}
  {\bibinfo  {journal} {Science}\ }\textbf {\bibinfo {volume} {343}},\ \bibinfo
  {pages} {393} (\bibinfo {year} {2014})}\BibitemShut {NoStop}%
\bibitem [{\citenamefont {Neto}\ \emph {et~al.}(2015)\citenamefont {Neto} \emph
  {et~al.}}]{SilvaNeto15a}%
  \BibitemOpen
  \bibfield  {author} {\bibinfo {author} {\bibfnamefont {E.~H. D.~S.}\
  \bibnamefont {Neto}} \emph {et~al.},\ }\href@noop {} {\bibfield  {journal}
  {\bibinfo  {journal} {Science}\ }\textbf {\bibinfo {volume} {347}},\ \bibinfo
  {pages} {282} (\bibinfo {year} {2015})}\BibitemShut {NoStop}%
\bibitem [{\citenamefont {Wu}\ \emph {et~al.}(2013)\citenamefont {Wu} \emph
  {et~al.}}]{Wu13a}%
  \BibitemOpen
  \bibfield  {author} {\bibinfo {author} {\bibfnamefont {T.}~\bibnamefont {Wu}}
  \emph {et~al.},\ }\href@noop {} {\bibfield  {journal} {\bibinfo  {journal}
  {Nature Communications}\ }\textbf {\bibinfo {volume} {4}},\ \bibinfo {pages}
  {2113} (\bibinfo {year} {2013})}\BibitemShut {NoStop}%
\bibitem [{\citenamefont {Andres}\ \emph {et~al.}(2005)\citenamefont {Andres},
  \citenamefont {Kartsovnik}, \citenamefont {Biberacher}, \citenamefont
  {Neumaier}, \citenamefont {Schuberth},\ and\ \citenamefont
  {M\protect{\"u}ller}}]{Andres05a}%
  \BibitemOpen
  \bibfield  {author} {\bibinfo {author} {\bibfnamefont {D.}~\bibnamefont
  {Andres}}, \bibinfo {author} {\bibfnamefont {M.~V.}\ \bibnamefont
  {Kartsovnik}}, \bibinfo {author} {\bibfnamefont {W.}~\bibnamefont
  {Biberacher}}, \bibinfo {author} {\bibfnamefont {K.}~\bibnamefont
  {Neumaier}}, \bibinfo {author} {\bibfnamefont {E.}~\bibnamefont {Schuberth}},
  \ and\ \bibinfo {author} {\bibfnamefont {H.}~\bibnamefont
  {M\protect{\"u}ller}},\ }\href@noop {} {\bibfield  {journal} {\bibinfo
  {journal} {Phys.\ Rev.\ B}\ }\textbf {\bibinfo {volume} {72}},\ \bibinfo
  {pages} {174513} (\bibinfo {year} {2005})}\BibitemShut {NoStop}%
\bibitem [{\citenamefont {Bangura}\ \emph {et~al.}(2005)\citenamefont
  {Bangura}, \citenamefont {Coldea}, \citenamefont {Singleton}, \citenamefont
  {Ardavan}, \citenamefont {Akutsu-Sato}, \citenamefont {Akutsu}, \citenamefont
  {Turner}, \citenamefont {Day}, \citenamefont {Yamamoto},\ and\ \citenamefont
  {Yakushi}}]{Bangura05a}%
  \BibitemOpen
  \bibfield  {author} {\bibinfo {author} {\bibfnamefont {A.~F.}\ \bibnamefont
  {Bangura}}, \bibinfo {author} {\bibfnamefont {A.~I.}\ \bibnamefont {Coldea}},
  \bibinfo {author} {\bibfnamefont {J.}~\bibnamefont {Singleton}}, \bibinfo
  {author} {\bibfnamefont {A.}~\bibnamefont {Ardavan}}, \bibinfo {author}
  {\bibfnamefont {A.}~\bibnamefont {Akutsu-Sato}}, \bibinfo {author}
  {\bibfnamefont {H.}~\bibnamefont {Akutsu}}, \bibinfo {author} {\bibfnamefont
  {S.~S.}\ \bibnamefont {Turner}}, \bibinfo {author} {\bibfnamefont
  {P.}~\bibnamefont {Day}}, \bibinfo {author} {\bibfnamefont {T.}~\bibnamefont
  {Yamamoto}}, \ and\ \bibinfo {author} {\bibfnamefont {K.}~\bibnamefont
  {Yakushi}},\ }\href@noop {} {\bibfield  {journal} {\bibinfo  {journal}
  {Phys.\ Rev.\ B}\ }\textbf {\bibinfo {volume} {72}},\ \bibinfo {pages}
  {014543} (\bibinfo {year} {2005})}\BibitemShut {NoStop}%
\bibitem [{\citenamefont {Tamura}\ \emph {et~al.}(2006)\citenamefont {Tamura},
  \citenamefont {Nakao},\ and\ \citenamefont {Kato}}]{Tamura06a}%
  \BibitemOpen
  \bibfield  {author} {\bibinfo {author} {\bibfnamefont {M.}~\bibnamefont
  {Tamura}}, \bibinfo {author} {\bibfnamefont {A.}~\bibnamefont {Nakao}}, \
  and\ \bibinfo {author} {\bibfnamefont {R.}~\bibnamefont {Kato}},\ }\href@noop
  {} {\bibfield  {journal} {\bibinfo  {journal} {J.\ Phys.\ Soc.\ Jpn.}\
  }\textbf {\bibinfo {volume} {75}},\ \bibinfo {pages} {093701} (\bibinfo
  {year} {2006})}\BibitemShut {NoStop}%
\bibitem [{\citenamefont {Shikama}\ \emph {et~al.}(2012)\citenamefont {Shikama}
  \emph {et~al.}}]{Shikama12a}%
  \BibitemOpen
  \bibfield  {author} {\bibinfo {author} {\bibfnamefont {T.}~\bibnamefont
  {Shikama}} \emph {et~al.},\ }\href@noop {} {\bibfield  {journal} {\bibinfo
  {journal} {Crystals}\ }\textbf {\bibinfo {volume} {2}},\ \bibinfo {pages}
  {1502} (\bibinfo {year} {2012})}\BibitemShut {NoStop}%
\bibitem [{\citenamefont {Li}\ \emph {et~al.}(2010)\citenamefont {Li},
  \citenamefont {Clay},\ and\ \citenamefont {S.Mazumdar}}]{Li10a}%
  \BibitemOpen
  \bibfield  {author} {\bibinfo {author} {\bibfnamefont {H.}~\bibnamefont
  {Li}}, \bibinfo {author} {\bibfnamefont {R.~T.}\ \bibnamefont {Clay}}, \ and\
  \bibinfo {author} {\bibnamefont {S.Mazumdar}},\ }\href@noop {} {\bibfield
  {journal} {\bibinfo  {journal} {J. Phys.: Condens. Matter}\ }\textbf
  {\bibinfo {volume} {22}},\ \bibinfo {pages} {272201} (\bibinfo {year}
  {2010})}\BibitemShut {NoStop}%
\bibitem [{\citenamefont {Dayal}\ \emph {et~al.}(2011)\citenamefont {Dayal},
  \citenamefont {Clay}, \citenamefont {Li},\ and\ \citenamefont
  {Mazumdar}}]{Dayal11a}%
  \BibitemOpen
  \bibfield  {author} {\bibinfo {author} {\bibfnamefont {S.}~\bibnamefont
  {Dayal}}, \bibinfo {author} {\bibfnamefont {R.~T.}\ \bibnamefont {Clay}},
  \bibinfo {author} {\bibfnamefont {H.}~\bibnamefont {Li}}, \ and\ \bibinfo
  {author} {\bibfnamefont {S.}~\bibnamefont {Mazumdar}},\ }\href@noop {}
  {\bibfield  {journal} {\bibinfo  {journal} {Phys.\ Rev.\ B}\ }\textbf
  {\bibinfo {volume} {83}},\ \bibinfo {pages} {245106} (\bibinfo {year}
  {2011})}\BibitemShut {NoStop}%
\bibitem [{\citenamefont {Moulopoulos}\ and\ \citenamefont
  {Ashcroft}(1992)}]{Moulopoulos92a}%
  \BibitemOpen
  \bibfield  {author} {\bibinfo {author} {\bibfnamefont {K.}~\bibnamefont
  {Moulopoulos}}\ and\ \bibinfo {author} {\bibfnamefont {N.~W.}\ \bibnamefont
  {Ashcroft}},\ }\href@noop {} {\bibfield  {journal} {\bibinfo  {journal}
  {Phys.\ Rev.\ Lett.}\ }\textbf {\bibinfo {volume} {69}},\ \bibinfo {pages}
  {2555} (\bibinfo {year} {1992})}\BibitemShut {NoStop}%
\bibitem [{\citenamefont {Gomi}\ \emph {et~al.}(2010)\citenamefont {Gomi},
  \citenamefont {Imai}, \citenamefont {Takahashi},\ and\ \citenamefont
  {Aihara}}]{Gomi10a}%
  \BibitemOpen
  \bibfield  {author} {\bibinfo {author} {\bibfnamefont {H.}~\bibnamefont
  {Gomi}}, \bibinfo {author} {\bibfnamefont {T.}~\bibnamefont {Imai}}, \bibinfo
  {author} {\bibfnamefont {A.}~\bibnamefont {Takahashi}}, \ and\ \bibinfo
  {author} {\bibfnamefont {M.}~\bibnamefont {Aihara}},\ }\href@noop {}
  {\bibfield  {journal} {\bibinfo  {journal} {Phys. Rev. B}\ }\textbf {\bibinfo
  {volume} {82}},\ \bibinfo {pages} {035101} (\bibinfo {year}
  {2010})}\BibitemShut {NoStop}%
\bibitem [{\citenamefont {Naka}\ and\ \citenamefont
  {Ishihara}(2010)}]{Naka10a}%
  \BibitemOpen
  \bibfield  {author} {\bibinfo {author} {\bibfnamefont {M.}~\bibnamefont
  {Naka}}\ and\ \bibinfo {author} {\bibfnamefont {S.}~\bibnamefont
  {Ishihara}},\ }\href@noop {} {\bibfield  {journal} {\bibinfo  {journal} {J.\
  Phys.\ Soc.\ Jpn.}\ }\textbf {\bibinfo {volume} {79}},\ \bibinfo {pages}
  {063707} (\bibinfo {year} {2010})}\BibitemShut {NoStop}%
\bibitem [{\citenamefont {Hotta}(2010)}]{Hotta10a}%
  \BibitemOpen
  \bibfield  {author} {\bibinfo {author} {\bibfnamefont {C.}~\bibnamefont
  {Hotta}},\ }\href@noop {} {\bibfield  {journal} {\bibinfo  {journal} {Phys.
  Rev. B}\ }\textbf {\bibinfo {volume} {82}},\ \bibinfo {pages} {241104}
  (\bibinfo {year} {2010})}\BibitemShut {NoStop}%
\bibitem [{\citenamefont {Abdel-Jawad}\ \emph {et~al.}(2010)\citenamefont
  {Abdel-Jawad}, \citenamefont {Terasaki}, \citenamefont {Sasaki},
  \citenamefont {Yoneyama}, \citenamefont {Kobayashi}, \citenamefont {Uesu},\
  and\ \citenamefont {Hotta}}]{Abdel-Jawad10a}%
  \BibitemOpen
  \bibfield  {author} {\bibinfo {author} {\bibfnamefont {M.}~\bibnamefont
  {Abdel-Jawad}}, \bibinfo {author} {\bibfnamefont {I.}~\bibnamefont
  {Terasaki}}, \bibinfo {author} {\bibfnamefont {T.}~\bibnamefont {Sasaki}},
  \bibinfo {author} {\bibfnamefont {N.}~\bibnamefont {Yoneyama}}, \bibinfo
  {author} {\bibfnamefont {N.}~\bibnamefont {Kobayashi}}, \bibinfo {author}
  {\bibfnamefont {Y.}~\bibnamefont {Uesu}}, \ and\ \bibinfo {author}
  {\bibfnamefont {C.}~\bibnamefont {Hotta}},\ }\href@noop {} {\bibfield
  {journal} {\bibinfo  {journal} {Phys.\ Rev.\ B}\ }\textbf {\bibinfo {volume}
  {82}},\ \bibinfo {pages} {125119} (\bibinfo {year} {2010})}\BibitemShut
  {NoStop}%
\bibitem [{\citenamefont {Lunkenheimer}\ \emph {et~al.}(2012)\citenamefont
  {Lunkenheimer}, \citenamefont {M{\"u}ller}, \citenamefont {Krohns},
  \citenamefont {Schrettle}, \citenamefont {Loidl}, \citenamefont {Hartmann},
  \citenamefont {Rommel}, \citenamefont {de~Souza}, \citenamefont {Hotta},
  \citenamefont {Schlueter},\ and\ \citenamefont {Lang}}]{Lunkenheimer12a}%
  \BibitemOpen
  \bibfield  {author} {\bibinfo {author} {\bibfnamefont {P.}~\bibnamefont
  {Lunkenheimer}}, \bibinfo {author} {\bibfnamefont {J.}~\bibnamefont
  {M{\"u}ller}}, \bibinfo {author} {\bibfnamefont {S.}~\bibnamefont {Krohns}},
  \bibinfo {author} {\bibfnamefont {F.}~\bibnamefont {Schrettle}}, \bibinfo
  {author} {\bibfnamefont {A.}~\bibnamefont {Loidl}}, \bibinfo {author}
  {\bibfnamefont {B.}~\bibnamefont {Hartmann}}, \bibinfo {author}
  {\bibfnamefont {R.}~\bibnamefont {Rommel}}, \bibinfo {author} {\bibfnamefont
  {M.}~\bibnamefont {de~Souza}}, \bibinfo {author} {\bibfnamefont
  {C.}~\bibnamefont {Hotta}}, \bibinfo {author} {\bibfnamefont
  {J.}~\bibnamefont {Schlueter}}, \ and\ \bibinfo {author} {\bibfnamefont
  {M.}~\bibnamefont {Lang}},\ }\href@noop {} {\bibfield  {journal} {\bibinfo
  {journal} {Nat. Mater.}\ }\textbf {\bibinfo {volume} {11}},\ \bibinfo {pages}
  {755} (\bibinfo {year} {2012})}\BibitemShut {NoStop}%
\bibitem [{\citenamefont {Sedlmeier}\ \emph {et~al.}(2012)\citenamefont
  {Sedlmeier}, \citenamefont {Els{\"a}sser}, \citenamefont {Neubauer},
  \citenamefont {Beyer}, \citenamefont {Wu}, \citenamefont {Ivek},
  \citenamefont {Tomic}, \citenamefont {Schlueter}, ,\ and\ \citenamefont
  {Dressel}}]{Sedlmeier12a}%
  \BibitemOpen
  \bibfield  {author} {\bibinfo {author} {\bibfnamefont {K.}~\bibnamefont
  {Sedlmeier}}, \bibinfo {author} {\bibfnamefont {S.}~\bibnamefont
  {Els{\"a}sser}}, \bibinfo {author} {\bibfnamefont {D.}~\bibnamefont
  {Neubauer}}, \bibinfo {author} {\bibfnamefont {R.}~\bibnamefont {Beyer}},
  \bibinfo {author} {\bibfnamefont {D.}~\bibnamefont {Wu}}, \bibinfo {author}
  {\bibfnamefont {T.}~\bibnamefont {Ivek}}, \bibinfo {author} {\bibfnamefont
  {S.}~\bibnamefont {Tomic}}, \bibinfo {author} {\bibfnamefont {J.~A.}\
  \bibnamefont {Schlueter}}, , \ and\ \bibinfo {author} {\bibfnamefont
  {M.}~\bibnamefont {Dressel}},\ }\href@noop {} {\bibfield  {journal} {\bibinfo
   {journal} {Phys.\ Rev.\ B}\ }\textbf {\bibinfo {volume} {86}},\ \bibinfo
  {pages} {245103} (\bibinfo {year} {2012})}\BibitemShut {NoStop}%
\bibitem [{\citenamefont {Dressel}\ \emph {et~al.}(2016)\citenamefont
  {Dressel}, \citenamefont {Lazic}, \citenamefont {Pustogow}, \citenamefont
  {Zhukova}, \citenamefont {Gorshunov}, \citenamefont {Schlueter},
  \citenamefont {Milat}, \citenamefont {Gumhalter},\ and\ \citenamefont
  {Tomic}}]{Dressel16a}%
  \BibitemOpen
  \bibfield  {author} {\bibinfo {author} {\bibfnamefont {M.}~\bibnamefont
  {Dressel}}, \bibinfo {author} {\bibfnamefont {P.}~\bibnamefont {Lazic}},
  \bibinfo {author} {\bibfnamefont {A.}~\bibnamefont {Pustogow}}, \bibinfo
  {author} {\bibfnamefont {E.}~\bibnamefont {Zhukova}}, \bibinfo {author}
  {\bibfnamefont {B.}~\bibnamefont {Gorshunov}}, \bibinfo {author}
  {\bibfnamefont {J.~A.}\ \bibnamefont {Schlueter}}, \bibinfo {author}
  {\bibfnamefont {O.}~\bibnamefont {Milat}}, \bibinfo {author} {\bibfnamefont
  {B.}~\bibnamefont {Gumhalter}}, \ and\ \bibinfo {author} {\bibfnamefont
  {S.}~\bibnamefont {Tomic}},\ }\href@noop {} {\bibfield  {journal} {\bibinfo
  {journal} {Phys.\ Rev.\ B}\ }\textbf {\bibinfo {volume} {93}},\ \bibinfo
  {pages} {081201(R)} (\bibinfo {year} {2016})}\BibitemShut {NoStop}%
\bibitem [{\citenamefont {Girlando}\ \emph {et~al.}(2014)\citenamefont
  {Girlando}, \citenamefont {Masino}, \citenamefont {Schlueter}, \citenamefont
  {Drichko}, \citenamefont {Kaiser},\ and\ \citenamefont
  {Dressel}}]{Girlando14a}%
  \BibitemOpen
  \bibfield  {author} {\bibinfo {author} {\bibfnamefont {A.}~\bibnamefont
  {Girlando}}, \bibinfo {author} {\bibfnamefont {M.}~\bibnamefont {Masino}},
  \bibinfo {author} {\bibfnamefont {J.~A.}\ \bibnamefont {Schlueter}}, \bibinfo
  {author} {\bibfnamefont {N.}~\bibnamefont {Drichko}}, \bibinfo {author}
  {\bibfnamefont {S.}~\bibnamefont {Kaiser}}, \ and\ \bibinfo {author}
  {\bibfnamefont {M.}~\bibnamefont {Dressel}},\ }\href@noop {} {\bibfield
  {journal} {\bibinfo  {journal} {Phys.\ Rev.\ B}\ }\textbf {\bibinfo {volume}
  {89}},\ \bibinfo {pages} {174503} (\bibinfo {year} {2014})}\BibitemShut
  {NoStop}%
\bibitem [{\citenamefont {Yamamoto}\ \emph {et~al.}()\citenamefont {Yamamoto},
  \citenamefont {Matsushita}, \citenamefont {Nakazawa}, \citenamefont
  {K.Yakushi}, \citenamefont {Tamura},\ and\ \citenamefont {Kato}}]{Yamamoto}%
  \BibitemOpen
  \bibfield  {author} {\bibinfo {author} {\bibfnamefont {T.}~\bibnamefont
  {Yamamoto}}, \bibinfo {author} {\bibfnamefont {K.}~\bibnamefont
  {Matsushita}}, \bibinfo {author} {\bibfnamefont {Y.}~\bibnamefont
  {Nakazawa}}, \bibinfo {author} {\bibnamefont {K.Yakushi}}, \bibinfo {author}
  {\bibfnamefont {M.}~\bibnamefont {Tamura}}, \ and\ \bibinfo {author}
  {\bibfnamefont {R.}~\bibnamefont {Kato}},\ }\href@noop {} {}\bibinfo {note}
  {Unpublished}\BibitemShut {NoStop}%
\bibitem [{\citenamefont {Yamashita}\ \emph {et~al.}(2008)\citenamefont
  {Yamashita}, \citenamefont {Nakazawa}, \citenamefont {Oguni}, \citenamefont
  {Oshima}, \citenamefont {Nojiri}, \citenamefont {Shimizu}, \citenamefont
  {Miyagawa},\ and\ \citenamefont {Kanoda}}]{Yamashita08a}%
  \BibitemOpen
  \bibfield  {author} {\bibinfo {author} {\bibfnamefont {S.}~\bibnamefont
  {Yamashita}}, \bibinfo {author} {\bibfnamefont {Y.}~\bibnamefont {Nakazawa}},
  \bibinfo {author} {\bibfnamefont {M.}~\bibnamefont {Oguni}}, \bibinfo
  {author} {\bibfnamefont {Y.}~\bibnamefont {Oshima}}, \bibinfo {author}
  {\bibfnamefont {H.}~\bibnamefont {Nojiri}}, \bibinfo {author} {\bibfnamefont
  {Y.}~\bibnamefont {Shimizu}}, \bibinfo {author} {\bibfnamefont
  {K.}~\bibnamefont {Miyagawa}}, \ and\ \bibinfo {author} {\bibfnamefont
  {K.}~\bibnamefont {Kanoda}},\ }\href@noop {} {\bibfield  {journal} {\bibinfo
  {journal} {Nature Phys.}\ }\textbf {\bibinfo {volume} {4}},\ \bibinfo {pages}
  {459} (\bibinfo {year} {2008})}\BibitemShut {NoStop}%
\bibitem [{\citenamefont {Yamashita}\ \emph {et~al.}(2009)\citenamefont
  {Yamashita}, \citenamefont {Nakata}, \citenamefont {Kasahara}, \citenamefont
  {Sasaki}, \citenamefont {Yoneyama}, \citenamefont {Kobayashi}, \citenamefont
  {Fujimoto}, \citenamefont {Shibauchi},\ and\ \citenamefont
  {Matsuda}}]{Yamashita09a}%
  \BibitemOpen
  \bibfield  {author} {\bibinfo {author} {\bibfnamefont {M.}~\bibnamefont
  {Yamashita}}, \bibinfo {author} {\bibfnamefont {N.}~\bibnamefont {Nakata}},
  \bibinfo {author} {\bibfnamefont {Y.}~\bibnamefont {Kasahara}}, \bibinfo
  {author} {\bibfnamefont {T.}~\bibnamefont {Sasaki}}, \bibinfo {author}
  {\bibfnamefont {N.}~\bibnamefont {Yoneyama}}, \bibinfo {author}
  {\bibfnamefont {N.}~\bibnamefont {Kobayashi}}, \bibinfo {author}
  {\bibfnamefont {S.}~\bibnamefont {Fujimoto}}, \bibinfo {author}
  {\bibfnamefont {T.}~\bibnamefont {Shibauchi}}, \ and\ \bibinfo {author}
  {\bibfnamefont {Y.}~\bibnamefont {Matsuda}},\ }\href@noop {} {\bibfield
  {journal} {\bibinfo  {journal} {Nature Phys.}\ }\textbf {\bibinfo {volume}
  {5}},\ \bibinfo {pages} {44} (\bibinfo {year} {2009})}\BibitemShut {NoStop}%
\bibitem [{\citenamefont {Manna}\ \emph {et~al.}(2010)\citenamefont {Manna},
  \citenamefont {de~Souza}, \citenamefont {Bruhl}, \citenamefont {Schlueter},\
  and\ \citenamefont {Lang}}]{Manna10a}%
  \BibitemOpen
  \bibfield  {author} {\bibinfo {author} {\bibfnamefont {R.~S.}\ \bibnamefont
  {Manna}}, \bibinfo {author} {\bibfnamefont {M.}~\bibnamefont {de~Souza}},
  \bibinfo {author} {\bibfnamefont {A.}~\bibnamefont {Bruhl}}, \bibinfo
  {author} {\bibfnamefont {J.~A.}\ \bibnamefont {Schlueter}}, \ and\ \bibinfo
  {author} {\bibfnamefont {M.}~\bibnamefont {Lang}},\ }\href@noop {} {\bibfield
   {journal} {\bibinfo  {journal} {Phys.\ Rev.\ Lett.}\ }\textbf {\bibinfo
  {volume} {104}},\ \bibinfo {pages} {016403} (\bibinfo {year}
  {2010})}\BibitemShut {NoStop}%
\bibitem [{\citenamefont {Poirier}\ \emph {et~al.}(2012)\citenamefont
  {Poirier}, \citenamefont {Parent}, \citenamefont {Cote}, \citenamefont
  {Miyagawa}, \citenamefont {Kanoda},\ and\ \citenamefont
  {Shimizu}}]{Poirier12a}%
  \BibitemOpen
  \bibfield  {author} {\bibinfo {author} {\bibfnamefont {M.}~\bibnamefont
  {Poirier}}, \bibinfo {author} {\bibfnamefont {S.}~\bibnamefont {Parent}},
  \bibinfo {author} {\bibfnamefont {A.}~\bibnamefont {Cote}}, \bibinfo {author}
  {\bibfnamefont {K.}~\bibnamefont {Miyagawa}}, \bibinfo {author}
  {\bibfnamefont {K.}~\bibnamefont {Kanoda}}, \ and\ \bibinfo {author}
  {\bibfnamefont {Y.}~\bibnamefont {Shimizu}},\ }\href@noop {} {\bibfield
  {journal} {\bibinfo  {journal} {Phys.\ Rev.\ B}\ }\textbf {\bibinfo {volume}
  {85}},\ \bibinfo {pages} {134444} (\bibinfo {year} {2012})}\BibitemShut
  {NoStop}%
\bibitem [{\citenamefont {Drichko}\ \emph {et~al.}(2014)\citenamefont
  {Drichko}, \citenamefont {Beyer}, \citenamefont {Rose}, \citenamefont
  {Dressel}, \citenamefont {Schlueter}, \citenamefont {Turunova}, \citenamefont
  {Zhilyaeva},\ and\ \citenamefont {Lyubovskaya}}]{Drichko14a}%
  \BibitemOpen
  \bibfield  {author} {\bibinfo {author} {\bibfnamefont {N.}~\bibnamefont
  {Drichko}}, \bibinfo {author} {\bibfnamefont {R.}~\bibnamefont {Beyer}},
  \bibinfo {author} {\bibfnamefont {E.}~\bibnamefont {Rose}}, \bibinfo {author}
  {\bibfnamefont {M.}~\bibnamefont {Dressel}}, \bibinfo {author} {\bibfnamefont
  {J.~A.}\ \bibnamefont {Schlueter}}, \bibinfo {author} {\bibfnamefont {S.~A.}\
  \bibnamefont {Turunova}}, \bibinfo {author} {\bibfnamefont {E.~I.}\
  \bibnamefont {Zhilyaeva}}, \ and\ \bibinfo {author} {\bibfnamefont {R.~N.}\
  \bibnamefont {Lyubovskaya}},\ }\href@noop {} {\bibfield  {journal} {\bibinfo
  {journal} {Phys.\ Rev.\ B}\ }\textbf {\bibinfo {volume} {89}},\ \bibinfo
  {pages} {075133} (\bibinfo {year} {2014})}\BibitemShut {NoStop}%
\bibitem [{\citenamefont {Hashimoto}\ \emph {et~al.}(2015)\citenamefont
  {Hashimoto}, \citenamefont {Kobayashi}, \citenamefont {Okamura},
  \citenamefont {Taniguchi}, \citenamefont {Ikemoto}, \citenamefont {Moriwaki},
  \citenamefont {Iguchi}, \citenamefont {Naka}, \citenamefont {Ishihara},\ and\
  \citenamefont {Sasaki}}]{Hashimoto15a}%
  \BibitemOpen
  \bibfield  {author} {\bibinfo {author} {\bibfnamefont {K.}~\bibnamefont
  {Hashimoto}}, \bibinfo {author} {\bibfnamefont {R.}~\bibnamefont
  {Kobayashi}}, \bibinfo {author} {\bibfnamefont {H.}~\bibnamefont {Okamura}},
  \bibinfo {author} {\bibfnamefont {H.}~\bibnamefont {Taniguchi}}, \bibinfo
  {author} {\bibfnamefont {Y.}~\bibnamefont {Ikemoto}}, \bibinfo {author}
  {\bibfnamefont {T.}~\bibnamefont {Moriwaki}}, \bibinfo {author}
  {\bibfnamefont {S.}~\bibnamefont {Iguchi}}, \bibinfo {author} {\bibfnamefont
  {M.}~\bibnamefont {Naka}}, \bibinfo {author} {\bibfnamefont {S.}~\bibnamefont
  {Ishihara}}, \ and\ \bibinfo {author} {\bibfnamefont {T.}~\bibnamefont
  {Sasaki}},\ }\href@noop {} {\bibfield  {journal} {\bibinfo  {journal} {Phys.\
  Rev.\ B}\ }\textbf {\bibinfo {volume} {92}},\ \bibinfo {pages} {085149}
  (\bibinfo {year} {2015})}\BibitemShut {NoStop}%
\bibitem [{\citenamefont {Gomes}\ \emph {et~al.}()\citenamefont {Gomes},
  \citenamefont {Silva}, \citenamefont {Dutta}, \citenamefont {Clay},\ and\
  \citenamefont {Mazumdar}}]{Gomes15a}%
  \BibitemOpen
  \bibfield  {author} {\bibinfo {author} {\bibfnamefont {N.}~\bibnamefont
  {Gomes}}, \bibinfo {author} {\bibfnamefont {W.~W.~D.}\ \bibnamefont {Silva}},
  \bibinfo {author} {\bibfnamefont {T.}~\bibnamefont {Dutta}}, \bibinfo
  {author} {\bibfnamefont {R.~T.}\ \bibnamefont {Clay}}, \ and\ \bibinfo
  {author} {\bibfnamefont {S.}~\bibnamefont {Mazumdar}},\ }\href@noop {}
  {\enquote {\bibinfo {title} {Coulomb enhanced superconducting pair
  correlations in the frustrated quarter-filled band},}\ }\bibinfo {note}
  {Preprint http://arxiv.org/abs/1505.07496}\BibitemShut {NoStop}%
\bibitem [{\citenamefont {Clay}\ and\ \citenamefont
  {Mazumdar}(2005)}]{Clay05c}%
  \BibitemOpen
  \bibfield  {author} {\bibinfo {author} {\bibfnamefont {R.~T.}\ \bibnamefont
  {Clay}}\ and\ \bibinfo {author} {\bibfnamefont {S.}~\bibnamefont
  {Mazumdar}},\ }\href@noop {} {\bibfield  {journal} {\bibinfo  {journal}
  {Synth.\ Metals}\ }\textbf {\bibinfo {volume} {153}},\ \bibinfo {pages} {445}
  (\bibinfo {year} {2005})}\BibitemShut {NoStop}%
\bibitem [{\citenamefont {Gomi}\ \emph {et~al.}(2013)\citenamefont {Gomi},
  \citenamefont {Ikenaga}, \citenamefont {Hiragi}, \citenamefont {Segawa},
  \citenamefont {Takahashi}, \citenamefont {Inagaki},\ and\ \citenamefont
  {Aihara}}]{Gomi13a}%
  \BibitemOpen
  \bibfield  {author} {\bibinfo {author} {\bibfnamefont {H.}~\bibnamefont
  {Gomi}}, \bibinfo {author} {\bibfnamefont {M.}~\bibnamefont {Ikenaga}},
  \bibinfo {author} {\bibfnamefont {Y.}~\bibnamefont {Hiragi}}, \bibinfo
  {author} {\bibfnamefont {D.}~\bibnamefont {Segawa}}, \bibinfo {author}
  {\bibfnamefont {A.}~\bibnamefont {Takahashi}}, \bibinfo {author}
  {\bibfnamefont {T.~J.}\ \bibnamefont {Inagaki}}, \ and\ \bibinfo {author}
  {\bibfnamefont {M.}~\bibnamefont {Aihara}},\ }\href@noop {} {\bibfield
  {journal} {\bibinfo  {journal} {Phys. Rev. B}\ }\textbf {\bibinfo {volume}
  {87}},\ \bibinfo {pages} {195126} (\bibinfo {year} {2013})}\BibitemShut
  {NoStop}%
\bibitem [{\citenamefont {Gomi}\ \emph {et~al.}(2016)\citenamefont {Gomi},
  \citenamefont {Inagaki},\ and\ \citenamefont {Takahashi}}]{Gomi16a}%
  \BibitemOpen
  \bibfield  {author} {\bibinfo {author} {\bibfnamefont {H.}~\bibnamefont
  {Gomi}}, \bibinfo {author} {\bibfnamefont {T.~J.}\ \bibnamefont {Inagaki}}, \
  and\ \bibinfo {author} {\bibfnamefont {A.}~\bibnamefont {Takahashi}},\
  }\href@noop {} {\bibfield  {journal} {\bibinfo  {journal} {Phys. Rev. B}\
  }\textbf {\bibinfo {volume} {93}},\ \bibinfo {pages} {035105} (\bibinfo
  {year} {2016})}\BibitemShut {NoStop}%
\bibitem [{\citenamefont {Ferber}\ \emph {et~al.}(2014)\citenamefont {Ferber},
  \citenamefont {Foyevtsova}, \citenamefont {Jeschke},\ and\ \citenamefont
  {Valenti}}]{Ferber14a}%
  \BibitemOpen
  \bibfield  {author} {\bibinfo {author} {\bibfnamefont {J.}~\bibnamefont
  {Ferber}}, \bibinfo {author} {\bibfnamefont {K.}~\bibnamefont {Foyevtsova}},
  \bibinfo {author} {\bibfnamefont {H.~O.}\ \bibnamefont {Jeschke}}, \ and\
  \bibinfo {author} {\bibfnamefont {R.}~\bibnamefont {Valenti}},\ }\href@noop
  {} {\bibfield  {journal} {\bibinfo  {journal} {Phys.\ Rev.\ B}\ }\textbf
  {\bibinfo {volume} {89}},\ \bibinfo {pages} {205106} (\bibinfo {year}
  {2014})}\BibitemShut {NoStop}%
\bibitem [{\citenamefont {Mori}\ \emph {et~al.}(1999)\citenamefont {Mori},
  \citenamefont {Mori},\ and\ \citenamefont {Tanaka}}]{Mori99a}%
  \BibitemOpen
  \bibfield  {author} {\bibinfo {author} {\bibfnamefont {T.}~\bibnamefont
  {Mori}}, \bibinfo {author} {\bibfnamefont {H.}~\bibnamefont {Mori}}, \ and\
  \bibinfo {author} {\bibfnamefont {S.}~\bibnamefont {Tanaka}},\ }\href@noop {}
  {\bibfield  {journal} {\bibinfo  {journal} {Bull. Chem. Soc. Jpn.}\ }\textbf
  {\bibinfo {volume} {72}},\ \bibinfo {pages} {179} (\bibinfo {year}
  {1999})}\BibitemShut {NoStop}%
\bibitem [{\citenamefont {Oike}\ \emph {et~al.}(2015)\citenamefont {Oike},
  \citenamefont {Miyagawa}, \citenamefont {Taniguchi},\ and\ \citenamefont
  {Kanoda}}]{Oike15a}%
  \BibitemOpen
  \bibfield  {author} {\bibinfo {author} {\bibfnamefont {H.}~\bibnamefont
  {Oike}}, \bibinfo {author} {\bibfnamefont {K.}~\bibnamefont {Miyagawa}},
  \bibinfo {author} {\bibfnamefont {H.}~\bibnamefont {Taniguchi}}, \ and\
  \bibinfo {author} {\bibfnamefont {K.}~\bibnamefont {Kanoda}},\ }\href@noop {}
  {\bibfield  {journal} {\bibinfo  {journal} {Phys.\ Rev.\ Lett.}\ }\textbf
  {\bibinfo {volume} {114}},\ \bibinfo {pages} {067002} (\bibinfo {year}
  {2015})}\BibitemShut {NoStop}%
\bibitem [{\citenamefont {Kandpal}\ \emph {et~al.}(2009)\citenamefont
  {Kandpal}, \citenamefont {Opahle}, \citenamefont {Zhang}, \citenamefont
  {Jeschke},\ and\ \citenamefont {Valent\'\i}}]{Kandpal09a}%
  \BibitemOpen
  \bibfield  {author} {\bibinfo {author} {\bibfnamefont {H.~C.}\ \bibnamefont
  {Kandpal}}, \bibinfo {author} {\bibfnamefont {I.}~\bibnamefont {Opahle}},
  \bibinfo {author} {\bibfnamefont {Y.-Z.}\ \bibnamefont {Zhang}}, \bibinfo
  {author} {\bibfnamefont {H.~O.}\ \bibnamefont {Jeschke}}, \ and\ \bibinfo
  {author} {\bibfnamefont {R.}~\bibnamefont {Valent\'\i}},\ }\href@noop {}
  {\bibfield  {journal} {\bibinfo  {journal} {Phys.\ Rev.\ Lett.}\ }\textbf
  {\bibinfo {volume} {103}},\ \bibinfo {pages} {067004} (\bibinfo {year}
  {2009})}\BibitemShut {NoStop}%
\bibitem [{\citenamefont {Nakamura}\ \emph {et~al.}(2009)\citenamefont
  {Nakamura}, \citenamefont {Yoshimoto}, \citenamefont {Kosugi}, \citenamefont
  {Arita},\ and\ \citenamefont {Imada}}]{Nakamura09a}%
  \BibitemOpen
  \bibfield  {author} {\bibinfo {author} {\bibfnamefont {K.}~\bibnamefont
  {Nakamura}}, \bibinfo {author} {\bibfnamefont {Y.}~\bibnamefont {Yoshimoto}},
  \bibinfo {author} {\bibfnamefont {T.}~\bibnamefont {Kosugi}}, \bibinfo
  {author} {\bibfnamefont {R.}~\bibnamefont {Arita}}, \ and\ \bibinfo {author}
  {\bibfnamefont {M.}~\bibnamefont {Imada}},\ }\href@noop {} {\bibfield
  {journal} {\bibinfo  {journal} {J.\ Phys.\ Soc.\ Jpn.}\ }\textbf {\bibinfo
  {volume} {78}},\ \bibinfo {pages} {083710} (\bibinfo {year}
  {2009})}\BibitemShut {NoStop}%
\bibitem [{\citenamefont {Koretsune}\ and\ \citenamefont
  {Hotta}(2014)}]{Koretsune14a}%
  \BibitemOpen
  \bibfield  {author} {\bibinfo {author} {\bibfnamefont {T.}~\bibnamefont
  {Koretsune}}\ and\ \bibinfo {author} {\bibfnamefont {C.}~\bibnamefont
  {Hotta}},\ }\href@noop {} {\bibfield  {journal} {\bibinfo  {journal} {Phys.\
  Rev.\ B}\ }\textbf {\bibinfo {volume} {89}},\ \bibinfo {pages} {045102}
  (\bibinfo {year} {2014})}\BibitemShut {NoStop}%
\bibitem [{sm()}]{sm}%
  \BibitemOpen
  \href@noop {} {}\bibinfo {note} {See Supplemental Material at
  http://link.aps.org/ supplemental/xx.xxxx/ PhysRevB.xxx.xxxxxx for further
  details of calculations.}\BibitemShut {Stop}%
\bibitem [{\citenamefont {Kashima}\ and\ \citenamefont
  {Imada}(2001)}]{Kashima01b}%
  \BibitemOpen
  \bibfield  {author} {\bibinfo {author} {\bibfnamefont {T.}~\bibnamefont
  {Kashima}}\ and\ \bibinfo {author} {\bibfnamefont {M.}~\bibnamefont
  {Imada}},\ }\href@noop {} {\bibfield  {journal} {\bibinfo  {journal} {J.\
  Phys.\ Soc.\ Jpn.}\ }\textbf {\bibinfo {volume} {70}},\ \bibinfo {pages}
  {2287} (\bibinfo {year} {2001})}\BibitemShut {NoStop}%
\bibitem [{\citenamefont {Zhang}\ \emph {et~al.}(1997)\citenamefont {Zhang},
  \citenamefont {Carlson},\ and\ \citenamefont {Gubernatis}}]{Zhang97a}%
  \BibitemOpen
  \bibfield  {author} {\bibinfo {author} {\bibfnamefont {S.}~\bibnamefont
  {Zhang}}, \bibinfo {author} {\bibfnamefont {J.}~\bibnamefont {Carlson}}, \
  and\ \bibinfo {author} {\bibfnamefont {J.~E.}\ \bibnamefont {Gubernatis}},\
  }\href@noop {} {\bibfield  {journal} {\bibinfo  {journal} {Phys.\ Rev.\ B}\
  }\textbf {\bibinfo {volume} {55}},\ \bibinfo {pages} {7464} (\bibinfo {year}
  {1997})}\BibitemShut {NoStop}%
\bibitem [{\citenamefont {Mizusaki}\ and\ \citenamefont
  {Imada}(2004)}]{Mizusaki04a}%
  \BibitemOpen
  \bibfield  {author} {\bibinfo {author} {\bibfnamefont {T.}~\bibnamefont
  {Mizusaki}}\ and\ \bibinfo {author} {\bibfnamefont {M.}~\bibnamefont
  {Imada}},\ }\href@noop {} {\bibfield  {journal} {\bibinfo  {journal} {Phys.\
  Rev.\ B}\ }\textbf {\bibinfo {volume} {69}},\ \bibinfo {pages} {125110}
  (\bibinfo {year} {2004})}\BibitemShut {NoStop}%
\bibitem [{\citenamefont {Huang}\ \emph {et~al.}(2001)\citenamefont {Huang},
  \citenamefont {Lin},\ and\ \citenamefont {Gubernatis}}]{Huang01a}%
  \BibitemOpen
  \bibfield  {author} {\bibinfo {author} {\bibfnamefont {Z.~B.}\ \bibnamefont
  {Huang}}, \bibinfo {author} {\bibfnamefont {H.~Q.}\ \bibnamefont {Lin}}, \
  and\ \bibinfo {author} {\bibfnamefont {J.~E.}\ \bibnamefont {Gubernatis}},\
  }\href@noop {} {\bibfield  {journal} {\bibinfo  {journal} {Phys.\ Rev.\ B}\
  }\textbf {\bibinfo {volume} {64}},\ \bibinfo {pages} {205101} (\bibinfo
  {year} {2001})}\BibitemShut {NoStop}%
\bibitem [{\citenamefont {Maguer{'e}s}\ \emph {et~al.}(1996)\citenamefont
  {Maguer{'e}s}, \citenamefont {Ouhab}, \citenamefont {Corian}, \citenamefont
  {Gomez-Garcia},\ and\ \citenamefont {Delhaes}}]{Magueres96a}%
  \BibitemOpen
  \bibfield  {author} {\bibinfo {author} {\bibfnamefont {P.~L.}\ \bibnamefont
  {Maguer{'e}s}}, \bibinfo {author} {\bibfnamefont {L.}~\bibnamefont {Ouhab}},
  \bibinfo {author} {\bibfnamefont {N.}~\bibnamefont {Corian}}, \bibinfo
  {author} {\bibfnamefont {C.~J.}\ \bibnamefont {Gomez-Garcia}}, \ and\
  \bibinfo {author} {\bibfnamefont {P.}~\bibnamefont {Delhaes}},\ }\href@noop
  {} {\bibfield  {journal} {\bibinfo  {journal} {Solid St.\ Comm.}\ }\textbf
  {\bibinfo {volume} {97}},\ \bibinfo {pages} {27} (\bibinfo {year}
  {1996})}\BibitemShut {NoStop}%
\bibitem [{\citenamefont {Swietlik}\ \emph {et~al.}(2003)\citenamefont
  {Swietlik}, \citenamefont {Lapinski}, \citenamefont {Polomska}, \citenamefont
  {Ouahab},\ and\ \citenamefont {Guillevic}}]{Swietlik03a}%
  \BibitemOpen
  \bibfield  {author} {\bibinfo {author} {\bibfnamefont {R.}~\bibnamefont
  {Swietlik}}, \bibinfo {author} {\bibfnamefont {A.}~\bibnamefont {Lapinski}},
  \bibinfo {author} {\bibfnamefont {M.}~\bibnamefont {Polomska}}, \bibinfo
  {author} {\bibfnamefont {L.}~\bibnamefont {Ouahab}}, \ and\ \bibinfo {author}
  {\bibfnamefont {J.}~\bibnamefont {Guillevic}},\ }\href@noop {} {\bibfield
  {journal} {\bibinfo  {journal} {Synth. Metals}\ }\textbf {\bibinfo {volume}
  {133}},\ \bibinfo {pages} {273} (\bibinfo {year} {2003})}\BibitemShut
  {NoStop}%
\bibitem [{\citenamefont {Ota}\ \emph {et~al.}(2007)\citenamefont {Ota},
  \citenamefont {Ouahab}, \citenamefont {Golhen}, \citenamefont {Yoshida},
  \citenamefont {Maesato}, \citenamefont {Saito},\ and\ \citenamefont
  {Swietlik}}]{Ota07a}%
  \BibitemOpen
  \bibfield  {author} {\bibinfo {author} {\bibfnamefont {A.}~\bibnamefont
  {Ota}}, \bibinfo {author} {\bibfnamefont {L.}~\bibnamefont {Ouahab}},
  \bibinfo {author} {\bibfnamefont {S.}~\bibnamefont {Golhen}}, \bibinfo
  {author} {\bibfnamefont {Y.}~\bibnamefont {Yoshida}}, \bibinfo {author}
  {\bibfnamefont {M.}~\bibnamefont {Maesato}}, \bibinfo {author} {\bibfnamefont
  {G.}~\bibnamefont {Saito}}, \ and\ \bibinfo {author} {\bibfnamefont
  {R.}~\bibnamefont {Swietlik}},\ }\href@noop {} {\bibfield  {journal}
  {\bibinfo  {journal} {Chemistry of Materials}\ }\textbf {\bibinfo {volume}
  {19}},\ \bibinfo {pages} {2455} (\bibinfo {year} {2007})}\BibitemShut
  {NoStop}%
\bibitem [{\citenamefont {Lapinski}\ \emph {et~al.}(2013)\citenamefont
  {Lapinski}, \citenamefont {Swietlik}, \citenamefont {Ouahab},\ and\
  \citenamefont {Golhen}}]{Lapinski13a}%
  \BibitemOpen
  \bibfield  {author} {\bibinfo {author} {\bibfnamefont {A.}~\bibnamefont
  {Lapinski}}, \bibinfo {author} {\bibfnamefont {R.}~\bibnamefont {Swietlik}},
  \bibinfo {author} {\bibfnamefont {L.}~\bibnamefont {Ouahab}}, \ and\ \bibinfo
  {author} {\bibfnamefont {S.}~\bibnamefont {Golhen}},\ }\href@noop {}
  {\bibfield  {journal} {\bibinfo  {journal} {J. Phys. Chem. A}\ }\textbf
  {\bibinfo {volume} {117}},\ \bibinfo {pages} {5241} (\bibinfo {year}
  {2013})}\BibitemShut {NoStop}%
\bibitem [{\citenamefont {Micnas}\ \emph {et~al.}(1990)\citenamefont {Micnas},
  \citenamefont {Ranninger},\ and\ \citenamefont {Robaszkiewicz}}]{Micnas90a}%
  \BibitemOpen
  \bibfield  {author} {\bibinfo {author} {\bibfnamefont {R.}~\bibnamefont
  {Micnas}}, \bibinfo {author} {\bibfnamefont {J.}~\bibnamefont {Ranninger}}, \
  and\ \bibinfo {author} {\bibfnamefont {S.}~\bibnamefont {Robaszkiewicz}},\
  }\href@noop {} {\bibfield  {journal} {\bibinfo  {journal} {Rev.\ Mod.\
  Phys.}\ }\textbf {\bibinfo {volume} {62}},\ \bibinfo {pages} {113} (\bibinfo
  {year} {1990})}\BibitemShut {NoStop}%
\bibitem [{\citenamefont {Mazumdar}\ and\ \citenamefont
  {Clay}(2008)}]{Mazumdar08a}%
  \BibitemOpen
  \bibfield  {author} {\bibinfo {author} {\bibfnamefont {S.}~\bibnamefont
  {Mazumdar}}\ and\ \bibinfo {author} {\bibfnamefont {R.~T.}\ \bibnamefont
  {Clay}},\ }\href@noop {} {\bibfield  {journal} {\bibinfo  {journal} {Phys.\
  Rev.\ B}\ }\textbf {\bibinfo {volume} {77}},\ \bibinfo {pages} {180515(R)}
  (\bibinfo {year} {2008})}\BibitemShut {NoStop}%
\bibitem [{\citenamefont {Holstein}(1959)}]{Holstein59a}%
  \BibitemOpen
  \bibfield  {author} {\bibinfo {author} {\bibfnamefont {T.}~\bibnamefont
  {Holstein}},\ }\href@noop {} {\bibfield  {journal} {\bibinfo  {journal} {Ann.
  Phys.(N.Y.)}\ }\textbf {\bibinfo {volume} {8}},\ \bibinfo {pages} {325}
  (\bibinfo {year} {1959})}\BibitemShut {NoStop}%
\bibitem [{\citenamefont {Su}\ \emph {et~al.}(1979)\citenamefont {Su},
  \citenamefont {Schrieffer},\ and\ \citenamefont {Heeger}}]{Su79a}%
  \BibitemOpen
  \bibfield  {author} {\bibinfo {author} {\bibfnamefont {W.~P.}\ \bibnamefont
  {Su}}, \bibinfo {author} {\bibfnamefont {J.~R.}\ \bibnamefont {Schrieffer}},
  \ and\ \bibinfo {author} {\bibfnamefont {A.~J.}\ \bibnamefont {Heeger}},\
  }\href@noop {} {\bibfield  {journal} {\bibinfo  {journal} {Phys.\ Rev.\
  Lett.}\ }\textbf {\bibinfo {volume} {42}},\ \bibinfo {pages} {1698} (\bibinfo
  {year} {1979})}\BibitemShut {NoStop}%
\bibitem [{\citenamefont {Mazumdar}\ and\ \citenamefont
  {Dixit}(1983)}]{Mazumdar83b}%
  \BibitemOpen
  \bibfield  {author} {\bibinfo {author} {\bibfnamefont {S.}~\bibnamefont
  {Mazumdar}}\ and\ \bibinfo {author} {\bibfnamefont {S.~N.}\ \bibnamefont
  {Dixit}},\ }\href@noop {} {\bibfield  {journal} {\bibinfo  {journal} {Phys.\
  Rev.\ Lett.}\ }\textbf {\bibinfo {volume} {51}},\ \bibinfo {pages} {292}
  (\bibinfo {year} {1983})}\BibitemShut {NoStop}%
\bibitem [{\citenamefont {Hirsch}(1983)}]{Hirsch83b}%
  \BibitemOpen
  \bibfield  {author} {\bibinfo {author} {\bibfnamefont {J.~E.}\ \bibnamefont
  {Hirsch}},\ }\href@noop {} {\bibfield  {journal} {\bibinfo  {journal} {Phys.\
  Rev.\ Lett.}\ }\textbf {\bibinfo {volume} {51}},\ \bibinfo {pages} {296}
  (\bibinfo {year} {1983})}\BibitemShut {NoStop}%
\bibitem [{\citenamefont {Mazumdar}\ and\ \citenamefont
  {Campbell}(1985)}]{Mazumdar85a}%
  \BibitemOpen
  \bibfield  {author} {\bibinfo {author} {\bibfnamefont {S.}~\bibnamefont
  {Mazumdar}}\ and\ \bibinfo {author} {\bibfnamefont {D.~K.}\ \bibnamefont
  {Campbell}},\ }\href@noop {} {\bibfield  {journal} {\bibinfo  {journal}
  {Phys.\ Rev.\ Lett.}\ }\textbf {\bibinfo {volume} {55}},\ \bibinfo {pages}
  {2067} (\bibinfo {year} {1985})}\BibitemShut {NoStop}%
\bibitem [{\citenamefont {Hirsch}\ and\ \citenamefont
  {Fradkin}(1983)}]{Hirsch83a}%
  \BibitemOpen
  \bibfield  {author} {\bibinfo {author} {\bibfnamefont {J.~E.}\ \bibnamefont
  {Hirsch}}\ and\ \bibinfo {author} {\bibfnamefont {E.}~\bibnamefont
  {Fradkin}},\ }\href@noop {} {\bibfield  {journal} {\bibinfo  {journal}
  {Phys.\ Rev.\ B}\ }\textbf {\bibinfo {volume} {27}},\ \bibinfo {pages} {4302}
  (\bibinfo {year} {1983})}\BibitemShut {NoStop}%
\bibitem [{\citenamefont {Assaad}\ \emph {et~al.}(1996)\citenamefont {Assaad},
  \citenamefont {Imada},\ and\ \citenamefont {Scalapino}}]{Assaad96a}%
  \BibitemOpen
  \bibfield  {author} {\bibinfo {author} {\bibfnamefont {F.~F.}\ \bibnamefont
  {Assaad}}, \bibinfo {author} {\bibfnamefont {M.}~\bibnamefont {Imada}}, \
  and\ \bibinfo {author} {\bibfnamefont {D.~J.}\ \bibnamefont {Scalapino}},\
  }\href@noop {} {\bibfield  {journal} {\bibinfo  {journal} {Phys.\ Rev.\
  Lett.}\ }\textbf {\bibinfo {volume} {77}},\ \bibinfo {pages} {4592} (\bibinfo
  {year} {1996})}\BibitemShut {NoStop}%
\bibitem [{\citenamefont {Assaad}\ \emph {et~al.}(1997)\citenamefont {Assaad},
  \citenamefont {Imada},\ and\ \citenamefont {Scalapino}}]{Assaad97a}%
  \BibitemOpen
  \bibfield  {author} {\bibinfo {author} {\bibfnamefont {F.~F.}\ \bibnamefont
  {Assaad}}, \bibinfo {author} {\bibfnamefont {M.}~\bibnamefont {Imada}}, \
  and\ \bibinfo {author} {\bibfnamefont {D.~J.}\ \bibnamefont {Scalapino}},\
  }\href@noop {} {\bibfield  {journal} {\bibinfo  {journal} {Phys.\ Rev.\ B}\
  }\textbf {\bibinfo {volume} {56}},\ \bibinfo {pages} {15001} (\bibinfo {year}
  {1997})}\BibitemShut {NoStop}%
\bibitem [{\citenamefont {Assaad}\ and\ \citenamefont
  {Imada}(1998)}]{Assaad98a}%
  \BibitemOpen
  \bibfield  {author} {\bibinfo {author} {\bibfnamefont {F.~F.}\ \bibnamefont
  {Assaad}}\ and\ \bibinfo {author} {\bibfnamefont {M.}~\bibnamefont {Imada}},\
  }\href@noop {} {\bibfield  {journal} {\bibinfo  {journal} {Phys.\ Rev.\ B}\
  }\textbf {\bibinfo {volume} {58}},\ \bibinfo {pages} {1845} (\bibinfo {year}
  {1998})}\BibitemShut {NoStop}%
\bibitem [{\citenamefont {Clay}\ \emph {et~al.}(2003)\citenamefont {Clay},
  \citenamefont {Mazumdar},\ and\ \citenamefont {Campbell}}]{Clay03a}%
  \BibitemOpen
  \bibfield  {author} {\bibinfo {author} {\bibfnamefont {R.~T.}\ \bibnamefont
  {Clay}}, \bibinfo {author} {\bibfnamefont {S.}~\bibnamefont {Mazumdar}}, \
  and\ \bibinfo {author} {\bibfnamefont {D.~K.}\ \bibnamefont {Campbell}},\
  }\href@noop {} {\bibfield  {journal} {\bibinfo  {journal} {Phys.\ Rev.\ B}\
  }\textbf {\bibinfo {volume} {67}},\ \bibinfo {pages} {115121} (\bibinfo
  {year} {2003})}\BibitemShut {NoStop}%
\bibitem [{\citenamefont {Hague}\ \emph {et~al.}(2007)\citenamefont {Hague},
  \citenamefont {Kornilovitch}, \citenamefont {Samson},\ and\ \citenamefont
  {Alexandrov}}]{Hague07a}%
  \BibitemOpen
  \bibfield  {author} {\bibinfo {author} {\bibfnamefont {J.~P.}\ \bibnamefont
  {Hague}}, \bibinfo {author} {\bibfnamefont {P.~E.}\ \bibnamefont
  {Kornilovitch}}, \bibinfo {author} {\bibfnamefont {J.~H.}\ \bibnamefont
  {Samson}}, \ and\ \bibinfo {author} {\bibfnamefont {A.~S.}\ \bibnamefont
  {Alexandrov}},\ }\href@noop {} {\bibfield  {journal} {\bibinfo  {journal}
  {Phys.\ Rev.\ Lett.}\ }\textbf {\bibinfo {volume} {98}},\ \bibinfo {pages}
  {037002} (\bibinfo {year} {2007})}\BibitemShut {NoStop}%
\bibitem [{\citenamefont {Vidmar}\ and\ \citenamefont
  {Bonca}(2010)}]{Vidmar10a}%
  \BibitemOpen
  \bibfield  {author} {\bibinfo {author} {\bibfnamefont {L.}~\bibnamefont
  {Vidmar}}\ and\ \bibinfo {author} {\bibfnamefont {J.}~\bibnamefont {Bonca}},\
  }\href@noop {} {\bibfield  {journal} {\bibinfo  {journal} {Phys.\ Rev.\ B}\
  }\textbf {\bibinfo {volume} {82}},\ \bibinfo {pages} {125121} (\bibinfo
  {year} {2010})}\BibitemShut {NoStop}%
\bibitem [{\citenamefont {Soto}\ \emph {et~al.}(1995)\citenamefont {Soto},
  \citenamefont {Slichter}, \citenamefont {Kini}, \citenamefont {Wang},
  \citenamefont {Geiser},\ and\ \citenamefont {Williams}}]{deSoto95a}%
  \BibitemOpen
  \bibfield  {author} {\bibinfo {author} {\bibfnamefont {S.~M.~D.}\
  \bibnamefont {Soto}}, \bibinfo {author} {\bibfnamefont {C.~P.}\ \bibnamefont
  {Slichter}}, \bibinfo {author} {\bibfnamefont {A.~M.}\ \bibnamefont {Kini}},
  \bibinfo {author} {\bibfnamefont {H.~H.}\ \bibnamefont {Wang}}, \bibinfo
  {author} {\bibfnamefont {U.}~\bibnamefont {Geiser}}, \ and\ \bibinfo {author}
  {\bibfnamefont {J.~M.}\ \bibnamefont {Williams}},\ }\href@noop {} {\bibfield
  {journal} {\bibinfo  {journal} {Phys.\ Rev.\ B}\ }\textbf {\bibinfo {volume}
  {52}},\ \bibinfo {pages} {10364} (\bibinfo {year} {1995})}\BibitemShut
  {NoStop}%
\bibitem [{\citenamefont {Kanoda}\ \emph {et~al.}(1996)\citenamefont {Kanoda},
  \citenamefont {Miyagawa}, \citenamefont {Kawamoto},\ and\ \citenamefont
  {Nakazawa}}]{Kanoda96a}%
  \BibitemOpen
  \bibfield  {author} {\bibinfo {author} {\bibfnamefont {K.}~\bibnamefont
  {Kanoda}}, \bibinfo {author} {\bibfnamefont {K.}~\bibnamefont {Miyagawa}},
  \bibinfo {author} {\bibfnamefont {A.}~\bibnamefont {Kawamoto}}, \ and\
  \bibinfo {author} {\bibfnamefont {Y.}~\bibnamefont {Nakazawa}},\ }\href@noop
  {} {\bibfield  {journal} {\bibinfo  {journal} {Phys.\ Rev.\ B}\ }\textbf
  {\bibinfo {volume} {54}},\ \bibinfo {pages} {76} (\bibinfo {year}
  {1996})}\BibitemShut {NoStop}%
\bibitem [{\citenamefont {Miyagawa}\ \emph {et~al.}(2004)\citenamefont
  {Miyagawa}, \citenamefont {Kanoda},\ and\ \citenamefont
  {Kawamoto}}]{Miyagawa04a}%
  \BibitemOpen
  \bibfield  {author} {\bibinfo {author} {\bibfnamefont {K.}~\bibnamefont
  {Miyagawa}}, \bibinfo {author} {\bibfnamefont {K.}~\bibnamefont {Kanoda}}, \
  and\ \bibinfo {author} {\bibfnamefont {A.}~\bibnamefont {Kawamoto}},\
  }\href@noop {} {\bibfield  {journal} {\bibinfo  {journal} {Chem. Rev.}\
  }\textbf {\bibinfo {volume} {104}},\ \bibinfo {pages} {5635} (\bibinfo {year}
  {2004})}\BibitemShut {NoStop}%
\bibitem [{\citenamefont {Shimizu}\ \emph {et~al.}(2010)\citenamefont
  {Shimizu}, \citenamefont {Kasahara}, \citenamefont {Furuta}, \citenamefont
  {Miyagawa}, \citenamefont {Kanoda}, \citenamefont {Maesato},\ and\
  \citenamefont {Saito}}]{Shimizu10a}%
  \BibitemOpen
  \bibfield  {author} {\bibinfo {author} {\bibfnamefont {Y.}~\bibnamefont
  {Shimizu}}, \bibinfo {author} {\bibfnamefont {H.}~\bibnamefont {Kasahara}},
  \bibinfo {author} {\bibfnamefont {T.}~\bibnamefont {Furuta}}, \bibinfo
  {author} {\bibfnamefont {K.}~\bibnamefont {Miyagawa}}, \bibinfo {author}
  {\bibfnamefont {K.}~\bibnamefont {Kanoda}}, \bibinfo {author} {\bibfnamefont
  {M.}~\bibnamefont {Maesato}}, \ and\ \bibinfo {author} {\bibfnamefont
  {G.}~\bibnamefont {Saito}},\ }\href@noop {} {\bibfield  {journal} {\bibinfo
  {journal} {Phys. Rev. B}\ }\textbf {\bibinfo {volume} {81}},\ \bibinfo
  {pages} {224508} (\bibinfo {year} {2010})}\BibitemShut {NoStop}%
\bibitem [{\citenamefont {Elsinger}\ \emph {et~al.}(2000)\citenamefont
  {Elsinger}, \citenamefont {Wosnitza}, \citenamefont {Wanka}, \citenamefont
  {Hagel}, \citenamefont {Schweitzer},\ and\ \citenamefont
  {Strunz}}]{Elsinger00a}%
  \BibitemOpen
  \bibfield  {author} {\bibinfo {author} {\bibfnamefont {H.}~\bibnamefont
  {Elsinger}}, \bibinfo {author} {\bibfnamefont {J.}~\bibnamefont {Wosnitza}},
  \bibinfo {author} {\bibfnamefont {S.}~\bibnamefont {Wanka}}, \bibinfo
  {author} {\bibfnamefont {J.}~\bibnamefont {Hagel}}, \bibinfo {author}
  {\bibfnamefont {D.}~\bibnamefont {Schweitzer}}, \ and\ \bibinfo {author}
  {\bibfnamefont {W.}~\bibnamefont {Strunz}},\ }\href@noop {} {\bibfield
  {journal} {\bibinfo  {journal} {Phys. Rev. Lett.}\ }\textbf {\bibinfo
  {volume} {84}},\ \bibinfo {pages} {6098} (\bibinfo {year}
  {2000})}\BibitemShut {NoStop}%
\bibitem [{\citenamefont {\protect{M\"uller}}\ \emph
  {et~al.}(2002)\citenamefont {\protect{M\"uller}}, \citenamefont {Lang},
  \citenamefont {Helfrich}, \citenamefont {Steglich},\ and\ \citenamefont
  {Sasaki}}]{Muller02b}%
  \BibitemOpen
  \bibfield  {author} {\bibinfo {author} {\bibfnamefont {J.}~\bibnamefont
  {\protect{M\"uller}}}, \bibinfo {author} {\bibfnamefont {M.}~\bibnamefont
  {Lang}}, \bibinfo {author} {\bibfnamefont {R.}~\bibnamefont {Helfrich}},
  \bibinfo {author} {\bibfnamefont {F.}~\bibnamefont {Steglich}}, \ and\
  \bibinfo {author} {\bibfnamefont {T.}~\bibnamefont {Sasaki}},\ }\href@noop {}
  {\bibfield  {journal} {\bibinfo  {journal} {Phys.\ Rev.\ B}\ }\textbf
  {\bibinfo {volume} {65}},\ \bibinfo {pages} {140509} (\bibinfo {year}
  {2002})}\BibitemShut {NoStop}%
\bibitem [{\citenamefont {Taylor}\ \emph {et~al.}(2007)\citenamefont {Taylor},
  \citenamefont {Carrington},\ and\ \citenamefont {Schlueter}}]{Taylor07a}%
  \BibitemOpen
  \bibfield  {author} {\bibinfo {author} {\bibfnamefont {O.~J.}\ \bibnamefont
  {Taylor}}, \bibinfo {author} {\bibfnamefont {A.}~\bibnamefont {Carrington}},
  \ and\ \bibinfo {author} {\bibfnamefont {J.~A.}\ \bibnamefont {Schlueter}},\
  }\href@noop {} {\bibfield  {journal} {\bibinfo  {journal} {Phys.\ Rev.\
  Lett.}\ }\textbf {\bibinfo {volume} {99}},\ \bibinfo {pages} {057001}
  (\bibinfo {year} {2007})}\BibitemShut {NoStop}%
\bibitem [{\citenamefont {Milbradt}\ \emph {et~al.}(2013)\citenamefont
  {Milbradt}, \citenamefont {Bardin}, \citenamefont {Truncik}, \citenamefont
  {Huttema}, \citenamefont {Jacko}, \citenamefont {Burn}, \citenamefont {Lo},
  \citenamefont {Powell},\ and\ \citenamefont {Broun}}]{Milbradt13a}%
  \BibitemOpen
  \bibfield  {author} {\bibinfo {author} {\bibfnamefont {S.}~\bibnamefont
  {Milbradt}}, \bibinfo {author} {\bibfnamefont {A.~A.}\ \bibnamefont
  {Bardin}}, \bibinfo {author} {\bibfnamefont {C.~J.~S.}\ \bibnamefont
  {Truncik}}, \bibinfo {author} {\bibfnamefont {W.~A.}\ \bibnamefont
  {Huttema}}, \bibinfo {author} {\bibfnamefont {A.~C.}\ \bibnamefont {Jacko}},
  \bibinfo {author} {\bibfnamefont {P.~L.}\ \bibnamefont {Burn}}, \bibinfo
  {author} {\bibfnamefont {S.-C.}\ \bibnamefont {Lo}}, \bibinfo {author}
  {\bibfnamefont {B.~J.}\ \bibnamefont {Powell}}, \ and\ \bibinfo {author}
  {\bibfnamefont {D.~M.}\ \bibnamefont {Broun}},\ }\href@noop {} {\bibfield
  {journal} {\bibinfo  {journal} {Phys. Rev. B}\ }\textbf {\bibinfo {volume}
  {88}},\ \bibinfo {pages} {064501} (\bibinfo {year} {2013})}\BibitemShut
  {NoStop}%
\bibitem [{\citenamefont {Perunov}\ \emph {et~al.}(2012)\citenamefont
  {Perunov}, \citenamefont {Shevchun}, \citenamefont {Kushch},\ and\
  \citenamefont {Trunin}}]{Perunov12a}%
  \BibitemOpen
  \bibfield  {author} {\bibinfo {author} {\bibfnamefont {N.~V.}\ \bibnamefont
  {Perunov}}, \bibinfo {author} {\bibfnamefont {A.~F.}\ \bibnamefont
  {Shevchun}}, \bibinfo {author} {\bibfnamefont {N.~D.}\ \bibnamefont
  {Kushch}}, \ and\ \bibinfo {author} {\bibfnamefont {M.~R.}\ \bibnamefont
  {Trunin}},\ }\href@noop {} {\bibfield  {journal} {\bibinfo  {journal} {JETP
  Letters}\ }\textbf {\bibinfo {volume} {96}},\ \bibinfo {pages} {184}
  (\bibinfo {year} {2012})}\BibitemShut {NoStop}%
\bibitem [{\citenamefont {Arai}\ \emph {et~al.}(2000)\citenamefont {Arai},
  \citenamefont {Ichimura}, \citenamefont {Nomura}, \citenamefont {Takasaki},
  \citenamefont {Yamada}, \citenamefont {Nakatsuji},\ and\ \citenamefont
  {Anzai}}]{Arai00a}%
  \BibitemOpen
  \bibfield  {author} {\bibinfo {author} {\bibfnamefont {T.}~\bibnamefont
  {Arai}}, \bibinfo {author} {\bibfnamefont {K.}~\bibnamefont {Ichimura}},
  \bibinfo {author} {\bibfnamefont {K.}~\bibnamefont {Nomura}}, \bibinfo
  {author} {\bibfnamefont {S.}~\bibnamefont {Takasaki}}, \bibinfo {author}
  {\bibfnamefont {J.}~\bibnamefont {Yamada}}, \bibinfo {author} {\bibfnamefont
  {S.}~\bibnamefont {Nakatsuji}}, \ and\ \bibinfo {author} {\bibfnamefont
  {H.}~\bibnamefont {Anzai}},\ }\href@noop {} {\bibfield  {journal} {\bibinfo
  {journal} {Solid St.\ Comm.}\ }\textbf {\bibinfo {volume} {116}},\ \bibinfo
  {pages} {679} (\bibinfo {year} {2000})}\BibitemShut {NoStop}%
\bibitem [{\citenamefont {Arai}\ \emph {et~al.}(2001)\citenamefont {Arai},
  \citenamefont {Ichimura}, \citenamefont {Nomura}, \citenamefont {Takasaki},
  \citenamefont {Yamada}, \citenamefont {Nakatsuji},\ and\ \citenamefont
  {Anzai}}]{Arai01a}%
  \BibitemOpen
  \bibfield  {author} {\bibinfo {author} {\bibfnamefont {T.}~\bibnamefont
  {Arai}}, \bibinfo {author} {\bibfnamefont {K.}~\bibnamefont {Ichimura}},
  \bibinfo {author} {\bibfnamefont {K.}~\bibnamefont {Nomura}}, \bibinfo
  {author} {\bibfnamefont {S.}~\bibnamefont {Takasaki}}, \bibinfo {author}
  {\bibfnamefont {J.}~\bibnamefont {Yamada}}, \bibinfo {author} {\bibfnamefont
  {S.}~\bibnamefont {Nakatsuji}}, \ and\ \bibinfo {author} {\bibfnamefont
  {H.}~\bibnamefont {Anzai}},\ }\href@noop {} {\bibfield  {journal} {\bibinfo
  {journal} {Phys. Rev. B}\ }\textbf {\bibinfo {volume} {63}},\ \bibinfo
  {pages} {104518} (\bibinfo {year} {2001})}\BibitemShut {NoStop}%
\bibitem [{\citenamefont {Ichimura}\ \emph {et~al.}(2003)\citenamefont
  {Ichimura}, \citenamefont {Suzuki}, \citenamefont {Nomura},\ and\
  \citenamefont {Kawamoto}}]{Ichimura03a}%
  \BibitemOpen
  \bibfield  {author} {\bibinfo {author} {\bibfnamefont {K.}~\bibnamefont
  {Ichimura}}, \bibinfo {author} {\bibfnamefont {K.}~\bibnamefont {Suzuki}},
  \bibinfo {author} {\bibfnamefont {K.}~\bibnamefont {Nomura}}, \ and\ \bibinfo
  {author} {\bibfnamefont {A.}~\bibnamefont {Kawamoto}},\ }\href@noop {}
  {\bibfield  {journal} {\bibinfo  {journal} {Synth.\ Metals}\ }\textbf
  {\bibinfo {volume} {133-134}},\ \bibinfo {pages} {213} (\bibinfo {year}
  {2003})}\BibitemShut {NoStop}%
\bibitem [{\citenamefont {Ichimura}\ \emph {et~al.}(2005)\citenamefont
  {Ichimura}, \citenamefont {Higashi}, \citenamefont {Nomura},\ and\
  \citenamefont {Kawamoto}}]{Ichimura05a}%
  \BibitemOpen
  \bibfield  {author} {\bibinfo {author} {\bibfnamefont {K.}~\bibnamefont
  {Ichimura}}, \bibinfo {author} {\bibfnamefont {S.}~\bibnamefont {Higashi}},
  \bibinfo {author} {\bibfnamefont {K.}~\bibnamefont {Nomura}}, \ and\ \bibinfo
  {author} {\bibfnamefont {A.}~\bibnamefont {Kawamoto}},\ }\href@noop {}
  {\bibfield  {journal} {\bibinfo  {journal} {Synth.\ Metals}\ }\textbf
  {\bibinfo {volume} {153}},\ \bibinfo {pages} {409} (\bibinfo {year}
  {2005})}\BibitemShut {NoStop}%
\bibitem [{\citenamefont {Ichimura}\ \emph {et~al.}(2008)\citenamefont
  {Ichimura}, \citenamefont {Takami},\ and\ \citenamefont
  {Nomura}}]{Ichimura08a}%
  \BibitemOpen
  \bibfield  {author} {\bibinfo {author} {\bibfnamefont {K.}~\bibnamefont
  {Ichimura}}, \bibinfo {author} {\bibfnamefont {M.}~\bibnamefont {Takami}}, \
  and\ \bibinfo {author} {\bibfnamefont {K.}~\bibnamefont {Nomura}},\
  }\href@noop {} {\bibfield  {journal} {\bibinfo  {journal} {Journal of the
  Physical Society of Japan}\ }\textbf {\bibinfo {volume} {77}},\ \bibinfo
  {pages} {114707} (\bibinfo {year} {2008})}\BibitemShut {NoStop}%
\bibitem [{\citenamefont {Oka}\ \emph {et~al.}(2015)\citenamefont {Oka},
  \citenamefont {Nobukane}, \citenamefont {Matsunaga}, \citenamefont {Nomura},
  \citenamefont {Katono}, \citenamefont {Ichimura},\ and\ \citenamefont
  {Kawamoto}}]{Oka15a}%
  \BibitemOpen
  \bibfield  {author} {\bibinfo {author} {\bibfnamefont {Y.}~\bibnamefont
  {Oka}}, \bibinfo {author} {\bibfnamefont {H.}~\bibnamefont {Nobukane}},
  \bibinfo {author} {\bibfnamefont {N.}~\bibnamefont {Matsunaga}}, \bibinfo
  {author} {\bibfnamefont {K.}~\bibnamefont {Nomura}}, \bibinfo {author}
  {\bibfnamefont {K.}~\bibnamefont {Katono}}, \bibinfo {author} {\bibfnamefont
  {K.}~\bibnamefont {Ichimura}}, \ and\ \bibinfo {author} {\bibfnamefont
  {A.}~\bibnamefont {Kawamoto}},\ }\href@noop {} {\bibfield  {journal}
  {\bibinfo  {journal} {Journal of the Physical Society of Japan}\ }\textbf
  {\bibinfo {volume} {84}},\ \bibinfo {pages} {064713} (\bibinfo {year}
  {2015})}\BibitemShut {NoStop}%
\bibitem [{\citenamefont {Schrama}\ \emph {et~al.}(1999)\citenamefont
  {Schrama}, \citenamefont {Rzepniewski}, \citenamefont {Edwards},
  \citenamefont {Singleton}, \citenamefont {Ardavan}, \citenamefont {Kurmoo},\
  and\ \citenamefont {Day}}]{Schrama99a}%
  \BibitemOpen
  \bibfield  {author} {\bibinfo {author} {\bibfnamefont {J.~M.}\ \bibnamefont
  {Schrama}}, \bibinfo {author} {\bibfnamefont {E.}~\bibnamefont
  {Rzepniewski}}, \bibinfo {author} {\bibfnamefont {R.~S.}\ \bibnamefont
  {Edwards}}, \bibinfo {author} {\bibfnamefont {J.}~\bibnamefont {Singleton}},
  \bibinfo {author} {\bibfnamefont {A.}~\bibnamefont {Ardavan}}, \bibinfo
  {author} {\bibfnamefont {M.}~\bibnamefont {Kurmoo}}, \ and\ \bibinfo {author}
  {\bibfnamefont {P.}~\bibnamefont {Day}},\ }\href@noop {} {\bibfield
  {journal} {\bibinfo  {journal} {Phys. Rev. Lett.}\ }\textbf {\bibinfo
  {volume} {83}},\ \bibinfo {pages} {3041} (\bibinfo {year}
  {1999})}\BibitemShut {NoStop}%
\bibitem [{\citenamefont {Malone}\ \emph {et~al.}(2010)\citenamefont {Malone},
  \citenamefont {Taylor}, \citenamefont {Schlueter},\ and\ \citenamefont
  {Carrington}}]{Malone10a}%
  \BibitemOpen
  \bibfield  {author} {\bibinfo {author} {\bibfnamefont {L.}~\bibnamefont
  {Malone}}, \bibinfo {author} {\bibfnamefont {O.~J.}\ \bibnamefont {Taylor}},
  \bibinfo {author} {\bibfnamefont {J.~A.}\ \bibnamefont {Schlueter}}, \ and\
  \bibinfo {author} {\bibfnamefont {A.}~\bibnamefont {Carrington}},\
  }\href@noop {} {\bibfield  {journal} {\bibinfo  {journal} {Phys. Rev. B}\
  }\textbf {\bibinfo {volume} {82}},\ \bibinfo {pages} {014522} (\bibinfo
  {year} {2010})}\BibitemShut {NoStop}%
\bibitem [{\citenamefont {Izawa}\ \emph {et~al.}(2001)\citenamefont {Izawa},
  \citenamefont {Yamaguchi}, \citenamefont {Sasaki},\ and\ \citenamefont
  {Matsuda}}]{Izawa01a}%
  \BibitemOpen
  \bibfield  {author} {\bibinfo {author} {\bibfnamefont {K.}~\bibnamefont
  {Izawa}}, \bibinfo {author} {\bibfnamefont {H.}~\bibnamefont {Yamaguchi}},
  \bibinfo {author} {\bibfnamefont {T.}~\bibnamefont {Sasaki}}, \ and\ \bibinfo
  {author} {\bibfnamefont {Y.}~\bibnamefont {Matsuda}},\ }\href@noop {}
  {\bibfield  {journal} {\bibinfo  {journal} {Phys.\ Rev.\ Lett.}\ }\textbf
  {\bibinfo {volume} {88}},\ \bibinfo {pages} {027002} (\bibinfo {year}
  {2001})}\BibitemShut {NoStop}%
\end{thebibliography}
\end{document}